\documentclass[fleqn,usenatbib]{mnras}

\usepackage{newtxtext,newtxmath}

\usepackage[T1]{fontenc}

\DeclareRobustCommand{\VAN}[3]{#2}
\let\VANthebibliography\thebibliography
\def\thebibliography{\DeclareRobustCommand{\VAN}[3]{##3}\VANthebibliography}

\usepackage{graphicx}	
\usepackage{amsmath}	
\usepackage{amsbsy}
\usepackage{mathtools}
\usepackage{tabularx}
\usepackage{breqn}
\usepackage{xcolor} 

\newcommand{\LOS}{\boldsymbol{\hat{n}}}

\definecolor{pablo}{rgb}{0.0, 0.5, 0.0}
\definecolor{lorne}{rgb}{1.0, 0.13, 0.32}
\definecolor{ofer}{rgb}{0.70, 0.051, 0.816}
\definecolor{KN}{rgb}{0.70, 0.051, 0.816}

\title[Integrated Sachs-Wolfe maps for MICE]{Full sky Integrated Sachs-Wolfe maps for the MICE Grand Challenge lightcone simulation}

\author[K. Naidoo et al.]{
  Krishna Naidoo,$^{1,2}$\thanks{E-mail: \href{mailto:knaidoo@cft.edu.pl}{knaidoo@cft.edu.pl}}
  Pablo Fosalba,$^{3,4}$
  Lorne Whiteway$^{2}$
  and Ofer Lahav$^{2}$\\
	\\
	$^{1}$Center for Theoretical Physics, Polish Academy of Sciences, Al. Lotnik\'{o}w 32/46, 02-668 Warsaw, Poland\\
	$^{2}$Department of Physics \& Astronomy, University College London, Gower Street, London WC1E 6BT, UK\\
	$^{3}$Institut d'Estudis Espacials de Catalunya (IEEC), 08034 Barcelona, Spain\\
	$^{4}$Institute of Space Sciences (ICE, CSIC), Campus UAB, Carrer de Can Magrans, s/n, 08193 Barcelona, Spain\\
}

\date{Accepted XXX. Received YYY; in original form ZZZ}

\pubyear{2020}

\begin{document}
\label{firstpage}
\pagerange{\pageref{firstpage}--\pageref{lastpage}}
\maketitle

\begin{abstract}
  We present full-sky maps of the Integrated Sachs-Wolfe effect (ISW) for the MICE
  Grand Challenge lightcone simulation up to redshift $1.4$. The maps are constructed
  in the linear regime using spherical Bessel transforms. We compare
  and contrast this procedure against analytical approximations found in the literature.
  By computing the ISW in the linear regime we remove the substantial computing
  and storage resources required to calculate the non-linear Rees-Sciama effect.
  Since the linear ISW at low redshift $z\lesssim 1$, at large angular scales, and after matter-domination is $\sim10^{2}\,\times$ larger in $\Delta T /T$ this has a negligible impact on the maps produced and only becomes relevant on scales which are dominated by cosmic microwave background (CMB) anisotropies.
  The MICE simulation products have been extensively used for studies involving current
  and future galaxy surveys. The availability of these maps will allow MICE to be
  used for future galaxy and CMB cross-correlation studies,
  ISW reconstruction studies and ISW void-stacking studies probed by galaxy surveys
  such as DES, DESI, \emph{Euclid} and Rubin LSST. The pipeline developed in this study is provided as a public \textsc{Python} package \textsc{pyGenISW}. This could be used in future studies for constructing the ISW from existing and future simulation suites probing vast sets of cosmological parameters and models.
\end{abstract}





\section{Introduction}

The Integrated Sachs-Wolfe effect \citep[ISW;][]{Sachs1967}, caused by the evolution of
gravitational potentials in large scale structure (LSS), imprints features onto
the cosmic microwave background (CMB). The strength of these features is sensitive
to the underlying cosmological model, in particular the quantity and nature
of dark energy \citep{Crittenden1996}. However, the CMB is dominated by primordial anisotropies meaning
the ISW is only detectable by cross-correlating the CMB with tracers of LSS.
This has been performed on several galaxy surveys to constrain the standard cosmological model $\Lambda$ Cold Dark Matter ($\Lambda$CDM) and to test expanded/alternative models \citep{Peiris2000, Boughn2002, Fosalba2003, Scranton2003, Nolta2004, Afshordi2004, Corasaniti2005, Padmanabhan2005, giannantoni2006, Vielva2006, McEwen2007, giannantonio2008, Ho2008, Xia2009, Dupe2011, giannantonio2012, Tomotsugu2012, giannantonio2014, Barreiro2013, PlanckISW2014, Cabass2015, Ferraro2015, Nicola2016, PlanckISW2016, Stolzner2018, Hang2020}.

Future galaxy redshift surveys will generate enormous catalogues of the position and
redshift of galaxies providing a large observational data set with which we may conduct
cross-correlation studies of LSS with the CMB. In order to understand the observational
sensitivities and systematics of the ISW to cosmological parameters we need to be
able to construct ISW maps and corresponding galaxy mocks for a wide range of
cosmological parameters and models. At present this is computationally expensive
as accurate ISW maps require regular snapshots of the gravitational potential.
This calculation therefore requires some foresight and cannot be done ad-hoc after
the simulations have been run. For this reason only a handful of such simulations
exist \citep[see][]{Cai2010, Watson2014, Carbone2016, Adamek2020} using either the best
fit Planck cosmology or exploring a limited set of cosmological parameters/models.

This limitation is most striking for ISW void-stacking studies \citep{Granett2008, Papai2011, Nadathur2012, Flender2013, Hernandez2013, Illic2013, Granett2015, Kovacs2016, Nadathur2016, Cai2017, Kovacs2018, Kovacs2019} which, to compare to theoretical predictions, require the simulation of realistic galaxy catalogues with corresponding ISW maps. Some
of these studies have measured a $\gtrsim 2\sigma$ excess in the ISW from large voids, the source of which remains unclear, but due to the computational cost of running ISW simulations have been limited to comparisons to
the fiducial $\Lambda$CDM model \citep[frequently comparing to the Jubilee ISW
maps;][]{Watson2014}.

The discovery of a void along the line-of-sight (LOS) of the CMB Cold Spot (CS) anomaly \citep{Szapudi2015} led \citet{Kovacs2016} to speculate whether the anomaly was actually caused by the same anomalous excess found for the ISW of other large voids. \citet{Naidoo2016} considered a $\Lambda$CDM solution to the CS anomaly -- multiple voids along the LOS \citep[later discovered by][]{Mackenzie2017}. However, even in the most extreme of scenarios multiple voids were unable to explain the full CS profile. \citet{Nadathur2014} showed that the significance of the CS is low ($\sim 2\sigma$) and could be explained as a tail-end Gaussian fluctuation. The significance of the CS was placed into further question when the effect of masking was considered and found to enhance the CS significance. By removing this effect the CS was found to be significant at only $\sim1.9\sigma$ \citep{Naidoo2017}. If the CS is evidence of the same anomalous excess in the void ISW signal, it will be difficult to establish from measurements of the CS alone. Instead, studies will need to further explore the signals from voids in data, using new and larger data sets, and by comparing to ISW simulations.

To explore the ISW for a larger set of cosmological parameters
and models, simulated ISW maps need to be easier to produce (so that these maps can be produced ad-hoc for a large set of existing and future
$N$-body simulations, without pre-planning). To do this we construct the ISW in the linear regime, removing
the costly requirement for regular snapshots of the gravitational potential across cosmic time. This
comes at a small cost: we lose the non-linear ISW \citep[known as the Rees-Sciama
effect;][]{Rees1968}, but since the ISW is $\sim10^{2}$ times larger in temperature \citep{Nadathur2014} this effect is negligible for the scales of interest in observational cosmology (which, due to the low signal-to-noise of the measurement, are generally limited to spherical harmonic modes of $\ell<100$; \citealt{Hang2020}). For $\ell \gtrsim 500$ the Rees-Sciama effect starts to dominate \citep{Cai2009} and for $\ell\gtrsim 5000$ it is the dominant source of CMB anisotropies \citep{Seljak1996}.

The ISW is constructed for the MICE Grand Challenge lightcone simulation \citep{MICE2015}
using a spherical Bessel transform method \citep{Shapiro2012} conducted in spherical
polar coordinates -- a natural coordinate system for lightcone simulations provided
in spherical shells \citep{Fosalba2008}.

The paper is organised as follows: in Section \ref{method} we describe the MICE
simulation data used and the ISW construction methods implemented in this study;
in Section \ref{results} we compare the ISW maps and their statistics to each other
and theoretical expectations; lastly, in Section \ref{discussion} we discuss
the results and the relevance of the data and pipeline produced in this study for future
work.

\section{Method}
\label{method}

\subsection{Data: The MICE Density Field}

The MICE Grand Challenge lightcone simulation \citep{MICE2015, MICE22015, MICE32015, MICE42015, MICE52015} is a large $N$-body
simulation constructed with \textsc{gadget}-2 \citep{Springel2005}. The simulation was run
with a comoving box of length $3072h^{-1}{\rm Mpc}$, $4096^{3}$ dark matter particles,
with a particle-mesh grid of $4096$ used to calculate large-scale forces computed
with Fast Fourier Transforms. The simulation used a flat $\Lambda$CDM cosmological
model consistent with the best-fit WMAP 5-year data \citep{Dunkley2009} -- i.e.
$\Omega_{\rm m,0}=0.25$, $\Omega_{\Lambda, 0}=0.75$, $\Omega_{\rm b,0}=0.044$, $n_{\rm s}=0.95$,
$\sigma_{8}=0.8$ and $h=0.7$.

The simulation's density contrast field is provided in an `onion' configuration \citep{Fosalba2008},
where particles from the MICE lightcone have been binned onto 400 \textsc{HEALPix} maps \citep{Gorski2005}.
The redshift slices are thinnest at low redshift and the thicknesses of the slices are well below the smallest
scales of interest ($k_{\max}=0.1\,h{\rm Mpc^{-1}}$). The simulation is large enough that no box repetitions are required along the LOS up to redshift $z=1.4$. For this reason the analysis in this paper is limited to $z\leq 1.4$. Corresponding galaxy mock catalogues can be obtained from the online database \textsc{CosmoHub}\footnote{\href{https://cosmohub.pic.es/home}{https://cosmohub.pic.es/home}} \citep{Cosmohub1, Cosmohub2}.

\subsection{Theory: The Integrated Sachs-Wolfe Effect}

We work in the Newtonian gauge with no anisotropic stress, so that metric perturbations can be parameterised by a single perturbation variable $\Phi$, the gravitational potential. The ISW is the imprint on the CMB of the evolution of $\Phi$ in large scale structure (LSS). The effect alters the CMB temperature $T$ in the LOS direction $\LOS$ by
\begin{equation}
\frac{\Delta T_{\textnormal{ISW}(\LOS)}}{T} = \frac{2}{c^2} \int_{\eta_{\textnormal{LS}}}^{\eta_0} \Phi'(r \LOS, \eta) \ d\eta,
\end{equation}
\noindent where the intergral is over a photon path from $\textrm{LS}$ (= last scattering time) to $0$ (= the observer's time), $r = c  (\eta_0 - \eta)$ is the comoving distance on this path corresponding to conformal time $\eta$, $c$ is the speed of light, and $\Phi'$ is the partial derivative of $\Phi$ with respect to $\eta$ (holding comoving position fixed). Changing the integration variable to comoving distance $r$ yields
\begin{equation}
\label{isw_basic_eq}
\frac{\Delta T_{\textnormal{ISW}(\LOS)}}{T} = \frac{2}{c^3} \int_{0}^{r_{\textnormal{LS}}} \dot{\Phi}(r \LOS, \eta(r)) \ a(r) \ dr.
\end{equation}
\noindent Here $\eta(r) = \eta_0 - r/c$ and we have additionally changed the integrand to be the partial derivative $\dot{\Phi}$ of $\Phi$ with respect to time (holding comoving position fixed); this introduces the scale factor $a = 1/(1+z)$ where $z = z(r)$ is the redshift. The relationship between redshift and comoving distance can be approximated (during and after matter domination in $\Lambda$CDM) by
\begin{equation}
r(z) \simeq 3000\, h^{-1}{\rm Mpc}\int_{0}^{z} \frac{1}{\Omega_{\rm m, 0}(1+z')^{3} + \Omega_{\rm \Lambda, 0}} dz',
\end{equation}
where $\Omega_{\rm m,0}$ is the current matter density and $\Omega_{\Lambda, 0}$
the current dark energy density (relative to the critical density).

The non-linear contributions to the ISW (the Rees-Sciama effect) have been shown to
be subdominant ($\sim 10^{-2}\times$ the linear ISW in temperature) in $\Lambda$CDM \citep{Seljak1996,
	Cai2010, Nadathur2014}. Therefore the ISW can be approximated to within $\sim1\%$
using linear perturbation theory, which we discuss below.

\subsubsection{Linear Theory Approximation}

We seek to calculate Eq. \ref{isw_basic_eq} in the linear regime. We start with the Poisson equation:
\begin{equation}
\label{eq:poisson}
\nabla^{2}\Phi(\pmb{x}, t) = \frac{3}{2}H_{0}^{2}\Omega_{\rm m, 0}\frac{\delta(\pmb{x},t)}{a}.
\end{equation}
In what follows we work in the linear regime. Here, density perturbations can be separated:
\begin{equation}
\label{eq:separation_of_delta}
\delta(\pmb{x}, t) = D(t)\delta(\pmb{x}),
\end{equation}
where the linear growth factor $D(z)$ is defined to be
\begin{equation}
D(z) \propto H(z)\int_{0}^{a}\frac{1}{(a' H(a'))^{3}} da'
\end{equation}
with $D(0)=1$ and $H(z)$ the Hubble expansion rate at redshift $z$.
The latter may be approximated during and after matter domination as
\begin{equation}
H(z) \simeq H_{0} \left(\Omega_{\rm m,0}(1+z)^{3} + \Omega_{\rm \Lambda,0}\right)^{1/2},
\end{equation}
where $H_{0}$, the Hubble constant, is the present expansion rate. Since we assume flat curvature, the energy density for dark energy is given by $\Omega_{\rm \Lambda}=1-\Omega_{\rm m}$ at any epoch.

Let $\nabla^{-2}$ denote the formal inverse of the Laplace operator. Combining Eqs. \ref{eq:poisson} and \ref{eq:separation_of_delta} yields
\begin{equation}
\label{eq:antipoisson}
\Phi(\pmb{x}, t) = \frac{3}{2}H_{0}^{2}\Omega_{\rm m,0}\frac{D(t)}{a}\nabla^{-2}\delta(\pmb{x}),
\end{equation}
whence
\begin{equation}
\label{eq:antipoisson_deriv}
\dot{\Phi}(\pmb{x}, t) = \frac{3}{2}H_{0}^{2}\Omega_{\rm m,0} \ \frac{\partial}{\partial t}\left( \frac{D(t)}{a} \right) \ \nabla^{-2}\delta(\pmb{x}).
\end{equation}
But
\begin{equation}
\label{eq:growth_factor_deriv}
\frac{\partial}{\partial t}\left(\frac{D(t)}{a}\right) = \frac{D(t)}{a}H(t)\left[f(t)-1\right];
\end{equation}
here
\begin{align}
f &\equiv {\rm d} \ln D / {\rm d} \ln a \nonumber\\
& \approx \Omega_{\rm m}(z)^{0.55}
\end{align}
\citep{Peebles1980, ofer1991}, and
\begin{equation}
\Omega_{\rm m}(z) = \frac{\Omega_{\rm m,0} (1+z)^{3}}{\Omega_{\rm m,0} (1+z)^{3} + \Omega_{\rm \Lambda,0}}.
\end{equation}

Combining Eqs. \ref{eq:antipoisson}, \ref{eq:antipoisson_deriv}, and \ref{eq:growth_factor_deriv} yields
\begin{equation}
\dot{\Phi}(\pmb{x},t) = H(t)\,\left[f(t)-1\right]\,\Phi(\pmb{x}, t),
\end{equation}
which coupled with Eq. \ref{isw_basic_eq} gives the linear theory approximation for the ISW:
\begin{equation}
\label{eq:linearISW}
\frac{\Delta T_{\rm ISW}(\pmb{\hat{n}})}{T} = \frac{2}{c^{3}}\int^{r_{\rm LS}}_{0}H(t)\,[f(t)-1]\, \Phi(r\pmb{\hat{n}}, \eta(r))\,a(r)\, dr.
\end{equation}
Note when $f\simeq1$, a condition which is true in an Einstein-de Sitter universe (where $\Omega_{\rm m}=1$ and $\Omega_{\Lambda}=0$) and during matter domination in $\Lambda$CDM, $\Delta T_{\rm ISW}\simeq0$. In this scenario the ISW is dominated by the non-linear Rees-Sciama effect. For this reason the most significant contributions to the ISW are at low redshift during $\Lambda$ domination.

\subsubsection{Theoretical Angular Power Spectra}

The angular power spectra $C_{\ell}^{XY}$ for sources $X$ and $Y$ is
\begin{equation}
C_{\ell}^{XY} = \frac{2}{\upi}\int k^{2}\,P(k)\,I_{\ell}^{X}(k)\,I_{\ell}^{Y}(k)\,dk,
\end{equation}
where
\begin{equation}
I_{\ell}^{X}(k)=\int_{0}^{\infty}D(z)\,W^{X}(r, k)\,j_{\ell}(kr)\,dr,
\end{equation}
$j_{\ell}$ is the spherical Bessel function and $W^{X}(r, k)$ is a source-specific
window function. For the ISW the window function is given by
\begin{equation}
W^{I}(r, k)=\frac{3\Omega_{m, 0}H_{0}^{2}}{c^{3}k^{2}}\,H(z)\,\left[1-f(z)\right]\,{\rm Rect}(r;r_{\min}, r_{\max}),
\end{equation}
where `$I$' is used as a shorthand for ISW and $\rm Rect$ is the rectangular step
function defined as
\begin{equation}
{\rm Rect}(r;r_{\min},r_{\max})={\rm H_{step}}(r-r_{\min})-{\rm H_{step}}(r-r_{\max}),
\end{equation}
$\rm H_{step}$ is the heavyside step function, $r_{\min}=r(z_{\min})$ and
$r_{\max}=r(z_{\max})$. This is used to specify the redshift interval $z_{\min}\leq z\leq z_{\max}$
for the contribution to the ISW. The galaxy window function is given by
\begin{equation}
\label{eq:wg}
W^{G}(r) = b(r)\,\Theta(r),
\end{equation}
where $b(r)$ is a function describing the galaxy bias,
\begin{equation}
\Theta(r) = \frac{r^{2}\,n(r)}{\int x^{2}\,n(x)\,dx},
\end{equation}
and $n(r)$ is the galaxy redshift distribution. However, in this study we have direct
access to the true underlying density contrast field meaning (unlike data from a galaxy survey) the redshift distribution and bias modelling can be simply set to $n(r)=1$ and $b(r)=1$ in the interval $z_{\min}\leq z\leq z_{\max}$ and $0$ otherwise.
The window function reduces to
\begin{equation}
W^{\delta}(r) = \begin{dcases}
\frac{3r^{2}}{r^{3}_{\max}-r^{3}_{\min}}, &\text{for } z_{\min}\leq z\leq z_{\max},\\
0, &\text{otherwise,}
\end{dcases}
\end{equation}
where `$\delta$' is used as a shorthand for the density contrast field in the interval $z_{\min}\leq z\leq z_{\max}$. It is common to calculate this using the Limber approximation \citep{Limber1954, Afshordi2004}, replacing the spherical
Bessel function with a Dirac delta function $\delta_{\rm D}$ of the form
\begin{equation}
j_{\ell}(x) \simeq \sqrt{\frac{\upi}{2\ell+1}}\,\delta_{\rm D}\left(\ell+\frac{1}{2}-x\right),
\end{equation}
giving the following approximation for the angular power spectra,
\begin{equation}
C_{\ell}^{XY}\simeq\int\frac{D^{2}(z)}{r^{2}}\left[P(k)\,W^{X}(r, k)\,W^{Y}(r, k)\right]_{k=k_{\ell}}dr,
\end{equation}
where $k_{\ell} = (\ell+1/2)/r$.

\begin{figure*}
	\centering
	\includegraphics[width=\textwidth]{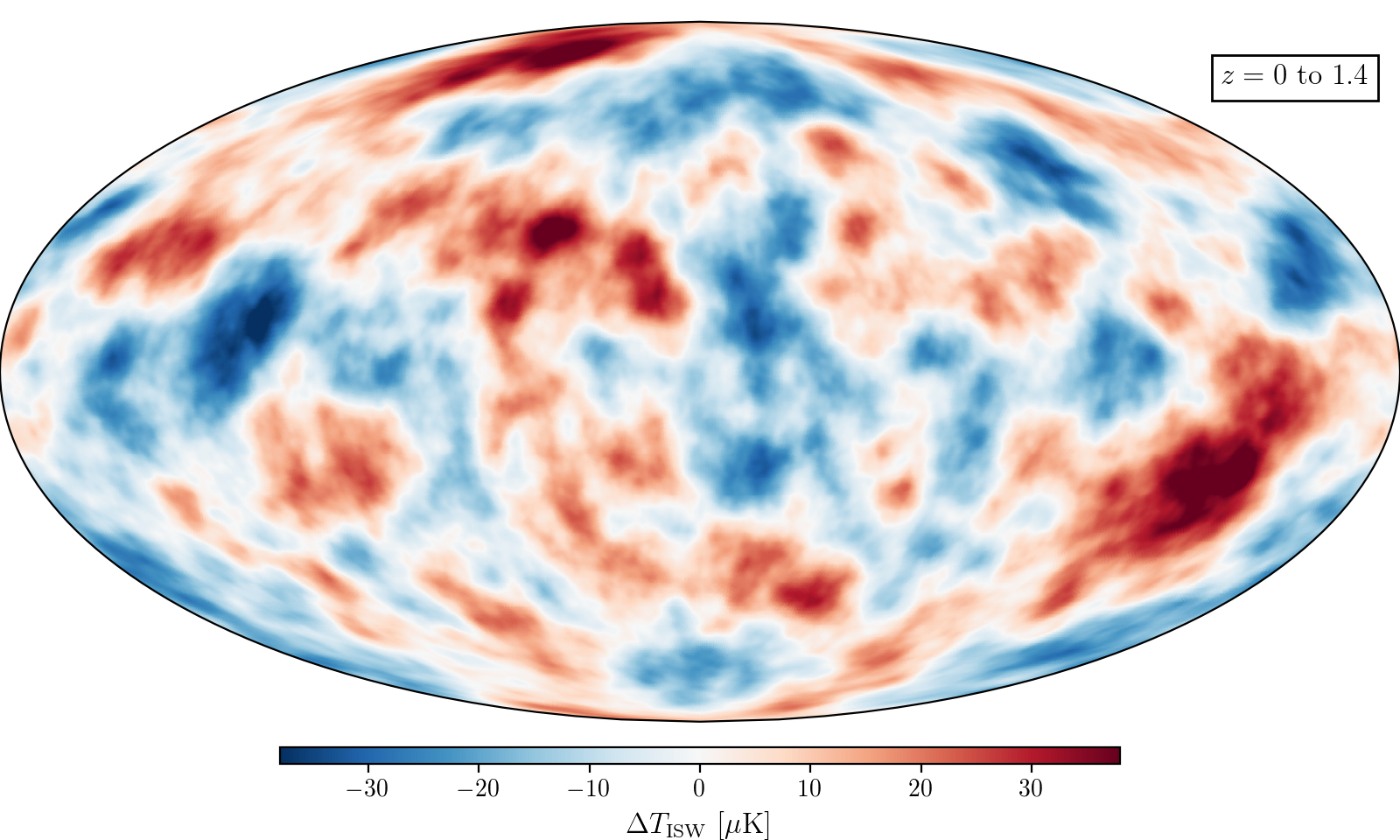}
	\caption{The ISW effect for the MICE simulation constructed for contributions in the range $0\leq z\leq 1.4$ using a SBT with normal boundary conditions. The map is shown with a blue-to-red diverging colormap indicating cold and hot features given in units $\mu {\rm K}$. The features of the map are quite broad in comparison to CMB maps (which typically have features on the scale of $\sim1\,{\rm deg}$) and are smaller in amplitude typically of the order of $\sim10\,\mu{\rm K}$ in the ISW compared to $\sim100\, \mu{\rm K}$ for the CMB.}
	\label{mice_isw}
\end{figure*}

\subsection{Techniques for Constructing Integrated Sachs-Wolfe Maps}

In this section we describe the numerical techniques used to construct ISW maps.
The focus of this study is the spherical Bessel transform (SBT) technique
\citep{Shapiro2012}, described in Section \ref{sbt}. We also compare this to a Spherical Harmonic scaling (which we call SHS) relation, described in Section \ref{Wiener}, and the \citet{Francis2010} technique, described
in Section \ref{Francis}.

\subsubsection{Spherical Bessel Transform Method}
\label{sbt}

A three dimensional field $\zeta$ described in spherical polar coordinates $r$ (radial axis),
$\theta$ and $\phi$ (latitude and longitude respectively) can be represented by its SBT
coefficients $\zeta_{\ell m n}$ as
\begin{equation}
\label{eq_backward_sbt}
\zeta(r,\theta,\phi) = \sum^{\infty}_{\ell=0}\sum^{\ell}_{m=-\ell}\sum_{n=1}^{\infty}\zeta_{\ell m n}\,R_{\ell n}(r)\,Y_{\ell m}(\theta, \phi),
\end{equation}
where
\begin{equation}
R_{\ell n}(r)=\frac{1}{\sqrt{N_{\ell n}}}j_{\ell}(k_{\ell n}r),
\end{equation}
$Y_{\ell m}$ are spherical harmonics, $k_{\ell n}=q_{\ell n}/r_{\max}$ and  $q_{\ell n}$ is
the locations of the $n^{\rm th}$ zero of $j_{\ell}(x)$ for normal boundary conditions
(i.e. $j_{\ell}(k_{\ell n}r_{\max})=0$) or $\partial_{x} j_{\ell}(x)$ for derivative boundary
conditions (i.e. $\partial_{x}j_{\ell}(k_{\ell n}r_{\max})=0$). Lastly, $N_{\ell n}$ is a normalisation
constant defined as
\begin{equation}
N_{\ell n}=\begin{dcases}
\frac{r_{\max}^{3}}{2}j_{\ell+1}^{2}(k_{\ell n}r_{\max}), &\text{normal boundary,}\\
\frac{r_{\max}^{3}}{2}\left(1 - \frac{\ell(\ell+1)}{k_{\ell n}^{2}r_{\max}^{2}}\right) j_{\ell}^{2}(k_{\ell n}r_{\max}), &\text{derivative boundary}.
\end{dcases}
\end{equation}
See \citet{Wang2009} for an overview of SBT and related transforms. The SBT basis functions are eigenfunctions of the Laplace operator on a ball-shaped domain. This will be crucial, as it is the Laplace operator in the Poisson equation that links the gravitational potential (which drives the ISW) to the overdensities (which are the observable quantities within the simulation).

The SBT coefficients are calculated from
\begin{equation}
\label{eq_forward_sbt}
\zeta_{\ell m n}=\int^{r_{\max}}_{0}\int^{\upi}_{0}\int^{2\upi}_{0}\zeta(r,\theta,\phi)\,R_{\ell n}(r)\,Y^{*}_{\ell m}(\theta, \phi)\, r^{2}\, \sin(\theta)\, d\phi\, d\theta\, dr,
\end{equation}
where we assume the field is defined up to a maximum radius $r_{\max}$.	We will be concerned with $\delta_{\ell mn}$, the SBT coefficients of the density contrast field $\delta$ $= \delta(\pmb{x})$ defined on three-dimensional space at $z=0$, and with $\Phi_{\ell mn}$, the SBT coefficients of the gravitational potential defined on the observer's lightcone.

Following \citet{Leistedt3DEX}
we calculate $\delta_{\ell mn}$ in two stages. The first step derives the spherical harmonics of
each slice $\delta^{i}$:
\begin{equation}
\delta_{\ell m}^{i}=\int_{0}^{\upi}\int_{0}^{2\upi} \frac{\delta^{i}(\theta,\phi)}{D(z_{\rm eff})}\,Y_{\ell m}^{*}(\theta, \phi)\,\sin(\theta)\,d\phi\,d\theta.
\end{equation}
Here we have corrected for the linear evolution of $\delta$ by dividing by the linear growth function at $z_{\rm eff}$ (the effective redshift of the slice). This ensures that the SBT coefficients of $\delta$ are taken at $z=0$.
The second step calculates the SBT coefficients from
\begin{equation}
\delta_{\ell m n} = \sum_{i} \delta_{\ell m}^{i}\int^{r^{i}_{\max}}_{r^{i}_{\min}}\,R_{\ell n}(r)\,r^{2}\,dr,
\end{equation}
where $r^{i}_{\min}$ and $r^{i}_{\max}$ are the minimum and maximum radius for the $i^{\rm th}$
slice.

Let $a^{I}_{\ell m}$ be the spherical harmonic coefficients for the ISW:
\begin{equation}
\label{eq:ISW_sph_harm_coeffs}
\frac{\Delta T_{\textnormal{ISW}}}{T} = \sum_{\ell=0}^{\infty}\sum_{m=-\ell}^{\ell} a^{I}_{\ell m} Y_{\ell m}(\theta,\phi).
\end{equation}
Equating spherical harmonic coefficients in Eqs. \ref{eq:linearISW} and \ref{eq:ISW_sph_harm_coeffs} gives
\begin{equation}
a^{I}_{\ell m} = \frac{2}{c^{3}}\int_{0}^{r_{\rm LS}}H(r)[f(r)-1] \sum_{n=1}^{\infty}\Phi_{\ell m n}R_{\ell n}(r)a(r)dr.
\end{equation}
The SBT basis function $R_{\ell n} Y_{\ell m}$ is an eigenfunction of the Laplace operator with eigenvalue $-{k_{\ell n}}^2$. Combining this with a SBT representation of the Poisson Eqn. \ref{eq:poisson} and applying Eqn. \ref{eq:separation_of_delta} yields
\begin{equation}
\Phi_{\ell mn} = -\frac{3}{2} H_0^2 \Omega_{m, 0} \frac{D(t)}{a} \frac{1}{{k_{\ell n}}^2} \delta_{\ell mn};
\end{equation}
combining the last two results gives our target expression for the ISW (due to \cite{Shapiro2012}):
\begin{equation}
a_{\ell m}^{I} = \frac{3 H_{0}^{2} \Omega_{m, 0}}{c^{3}}\sum_{n=1}^{n_{\max}}\frac{\delta_{\ell m n}}{{k_{\ell n}}^{2}}\int_{0}^{r_{\max}}D(r)H(r)\left[1-f(r)\right]R_{\ell n}(r)dr.
\end{equation}
In this study we compute only the SBT coefficients that correspond to Fourier modes in the range $k_{\rm F}\leq k\leq k_{\max}$ where $k_{\max}=0.1\,h{\rm Mpc^{-1}}$ and $k_{\rm F}$ is the fundamental frequency of the simulation ($k_{\rm F}=2\upi/L_{\rm box}$ where $L_{\rm box}$ is the length of the simulation box). This means in our analysis the SBT coefficients are computed for $\ell=2$ to $\ell_{\max}$ (ignoring the monopole and dipole components) and $n=1$ to $n_{\max}$, where $\ell_{\max} = \lfloor r_{\max}k_{\max}\rfloor$ and $n_{\max}=\lfloor\ell_{\max}/\upi\rfloor$ (the floor function ${\rm Floor}(x)=\lfloor x\rfloor$ ensures $l_{\max}$ and $n_{\max}$ are integers).

\begin{figure*}
	\centering
	\includegraphics[width=\textwidth]{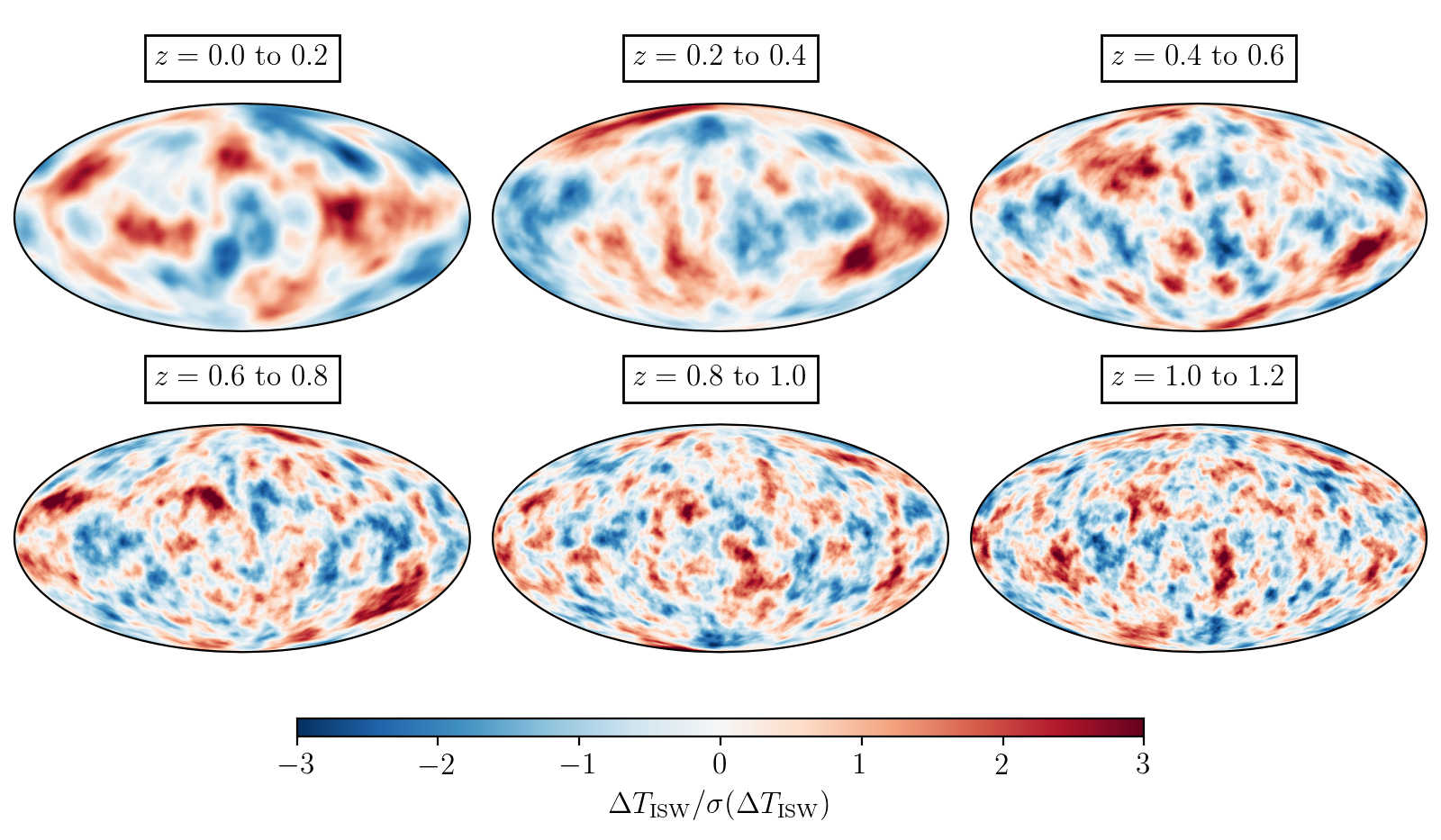}
	\caption{The ISW constructed using the SBT with normal boundary conditions shown for contributions in redshift slices of $\Delta z=0.2$ in the range $0\leq z\leq 1.2$. The maps are divided by their standard deviation to highlight the scale of the features in these different redshift slices. This change of resolution originates from the size of structures and the scale of homogeneity, at closer redshift the features are broad due to the size of nearby cosmic structures but at high redshift features are smaller as these cosmic structures are further away.
}
	\label{mice_isw_vs_z}
\end{figure*}

\subsubsection{Spherical Harmonic Scaling}
\label{Wiener}

A common alternative approach to constructing ISW maps from the density field is to use the SHS relation \citep{Manzotti2014, Muir2016},
\begin{equation}
a^{I}_{\ell m} = \frac{C_{\ell}^{I\delta}}{C_{\ell}^{\delta\delta}}a^{\delta}_{\ell m},
\end{equation}
where $C_{\ell}^{\delta\delta}$ is the auto-angular power spectrum for the density contrast, $C_{\ell}^{I\delta}$ the cross-angular power spectrum for the ISW-density contrast and $a^{\delta}_{\ell m}$ the spherical harmonic coefficients for the density field integrated over the redshift region of interest.

\begin{figure*}
	\centering
	\includegraphics[width=\textwidth]{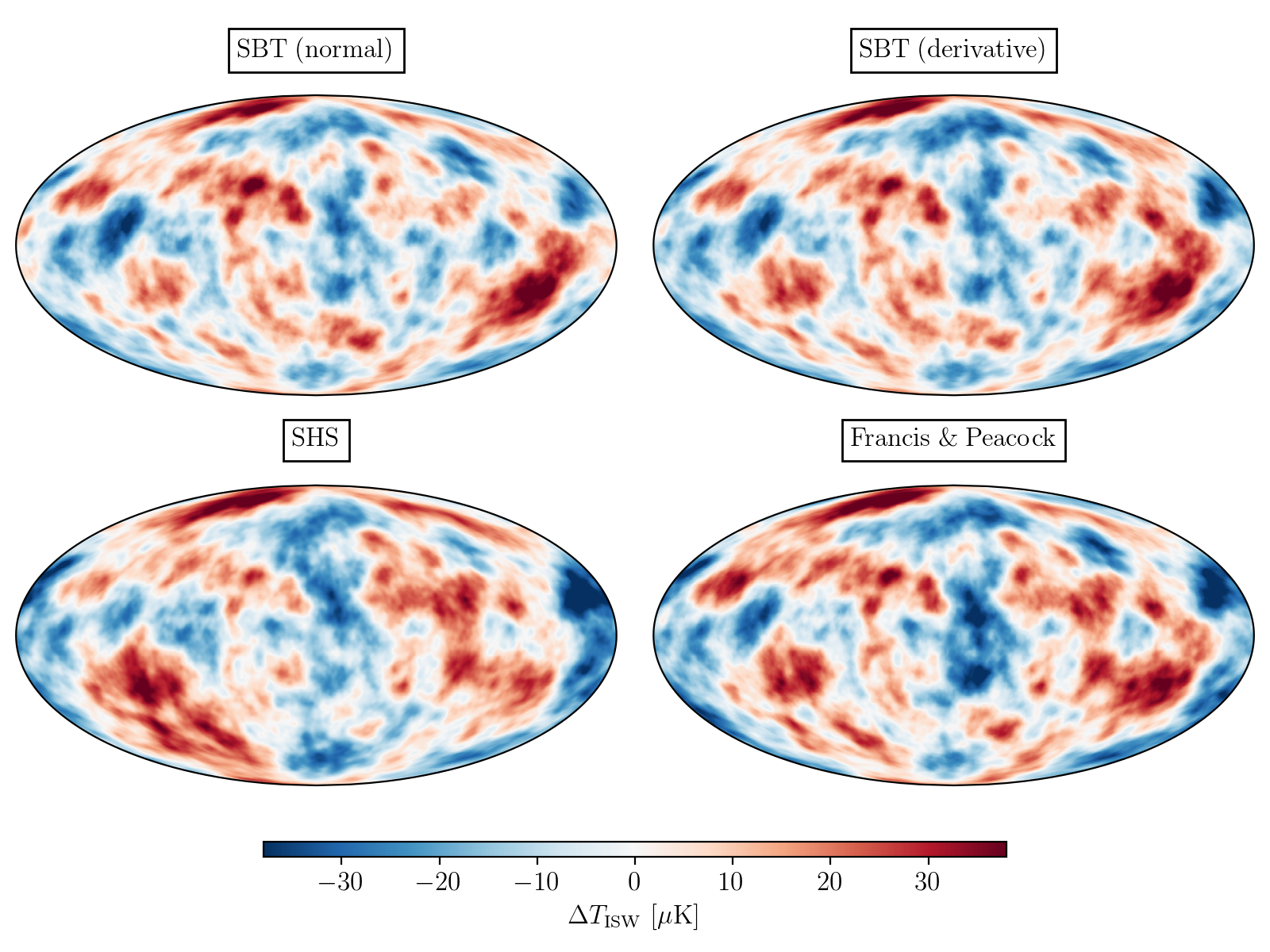}
	\caption{The ISW maps are shown for four methods for contributions from the redshift range $0\leq z\leq 1.4$. In the top left panel is the SBT (normal) map, in the top right panel is the SBT (derivative) map, in the bottom left panel is the SHS map and in the bottom right panel is the FP map. The maps are in fairly good agreement but the approximations on the bottom panels show variations in the amplitude of certain features while the SBT methods are almost identical.}
	\label{mice_comparisons_maps}
\end{figure*}

\begin{figure*}
	\centering
	\includegraphics[width=\textwidth]{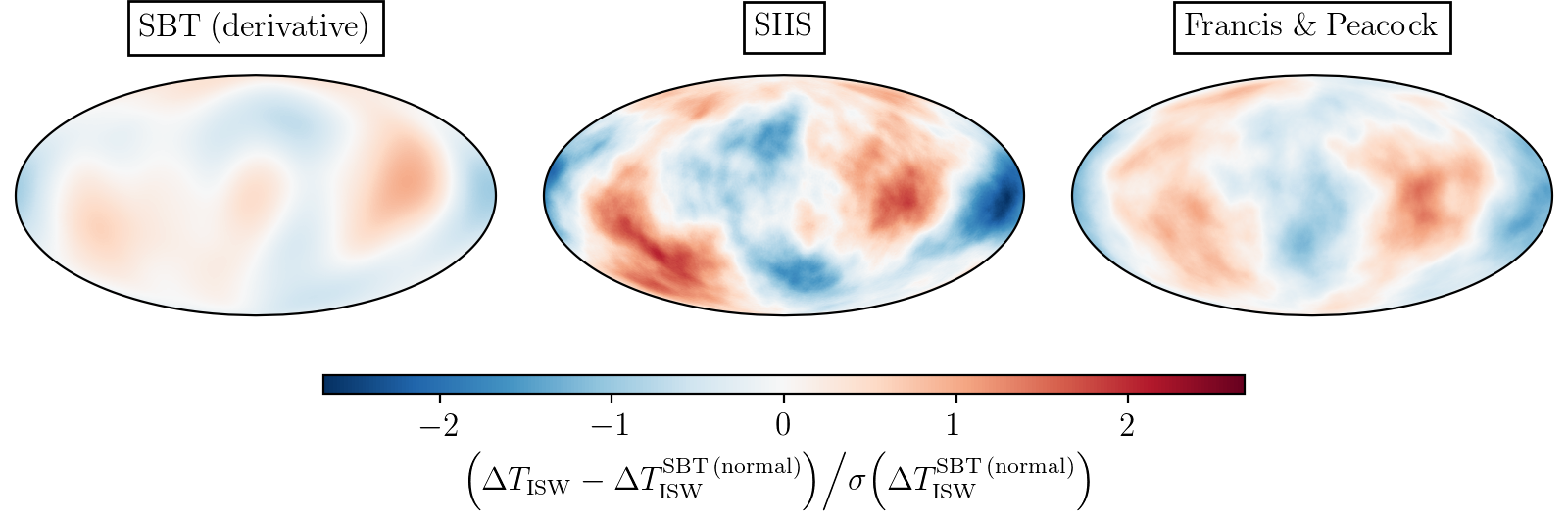}
	\caption{Differences between the SBT (normal) map and maps generated by the three other methods (normalised by the standard deviation of the SBT (normal) map) measured in the redshift range $0\leq z\leq 1.4$. From left to right are the difference for the SBT (derivative) map, for the SHS map, and for the FP map (all with respect to the SBT (normal) map). This shows only a small difference between the SBT methods (which differ only in their boundary condition) and shows a significant difference between the SHS and FP maps (which unlike the SBT methods are computed without any LOS information).}
	\label{mice_comparisons_dif}
\end{figure*}

\subsubsection{Francis \& Peacock Approximation}
\label{Francis}

\citet{Francis2010} derive an approximation for the ISW (now referred to as the FP approximation),
\begin{equation}
a_{\ell m}^{I}\simeq \frac{3\Omega_{\rm m,0} H_{0}^{2}}{\ell(\ell+1)c^{3}}\,H(z_{\rm eff}) \left[1-f(z_{\rm eff})\right]\,\Delta r \, r_{\rm eff}^{2}\,a_{\ell m}^{\delta},
\end{equation}
where $r_{\rm eff}$ is the effective comoving radius of the density contrast field,
\begin{equation}
r_{\rm eff}=\frac{3}{4}\left(\frac{r_{\max}^{4}-r_{\min}^{4}}{r_{\max}^{3}-r_{\min}^{3}}\right),
\end{equation}
$z_{\rm eff}$ is the redshift corresponding to $r_{\rm eff}$, $\Delta r=r_{\max}-r_{\min}$, $r_{\min}$ is the minimum comoving radius and $r_{\max}$ is the maximum comoving radius. This approximation is best suited for cases where $\Delta r$ is thin; for this reason with this approximation we use thin slices and then combine the maps.

\subsubsection{Software Pipeline}

The \textsc{Python} \citep{python3} package \textsc{pyGenISW} is made publicly available\footnote{\href{https://github.com/knaidoo29/pyGenISW}{https://github.com/knaidoo29/pyGenISW}} and can be used to construct ISW maps from data provided in spherical shells (given in \textsc{HEALPix} format) using the SBT method (which is the focus of this paper) as well as the alternative SHS approach and the FP approximation.

The package depends on \textsc{TheoryCL}\footnote{\href{https://github.com/knaidoo29/TheoryCL}{https://github.com/knaidoo29/TheoryCL}} which computes the linear growth functions and angular power spectra for the ISW and density contrast sources; \textsc{camb}\footnote{\href{https://camb.readthedocs.io/}{https://camb.readthedocs.io/}} \citep{Lewis1999} to compute the linear power spectrum, \textsc{healpy}\footnote{\href{https://healpy.readthedocs.io/}{https://healpy.readthedocs.io/}} \citep{Gorski2005, Zonca2019} for computing and manipulating maps and carrying out spherical harmonic operations; \textsc{SciPy} used for spherical Bessel related functions, integration functions and interpolation functions and \textsc{NumPy}\footnote{\href{https://numpy.org/}{https://numpy.org/}} \citep{Numpy2020}. Note, \textsc{pyGenISW} uses a mixture of \textsc{SciPy}\footnote{\href{https://www.scipy.org/}{https://www.scipy.org/}} \citep{Scipy2020} and its own iterative spherical Bessel root finding function to determine $q_{\ell n}$; this is because the current implementation in \textsc{SciPy} is unstable for large $\ell$ and $n$.

\begin{figure*}
	\centering
	\includegraphics[width=\textwidth]{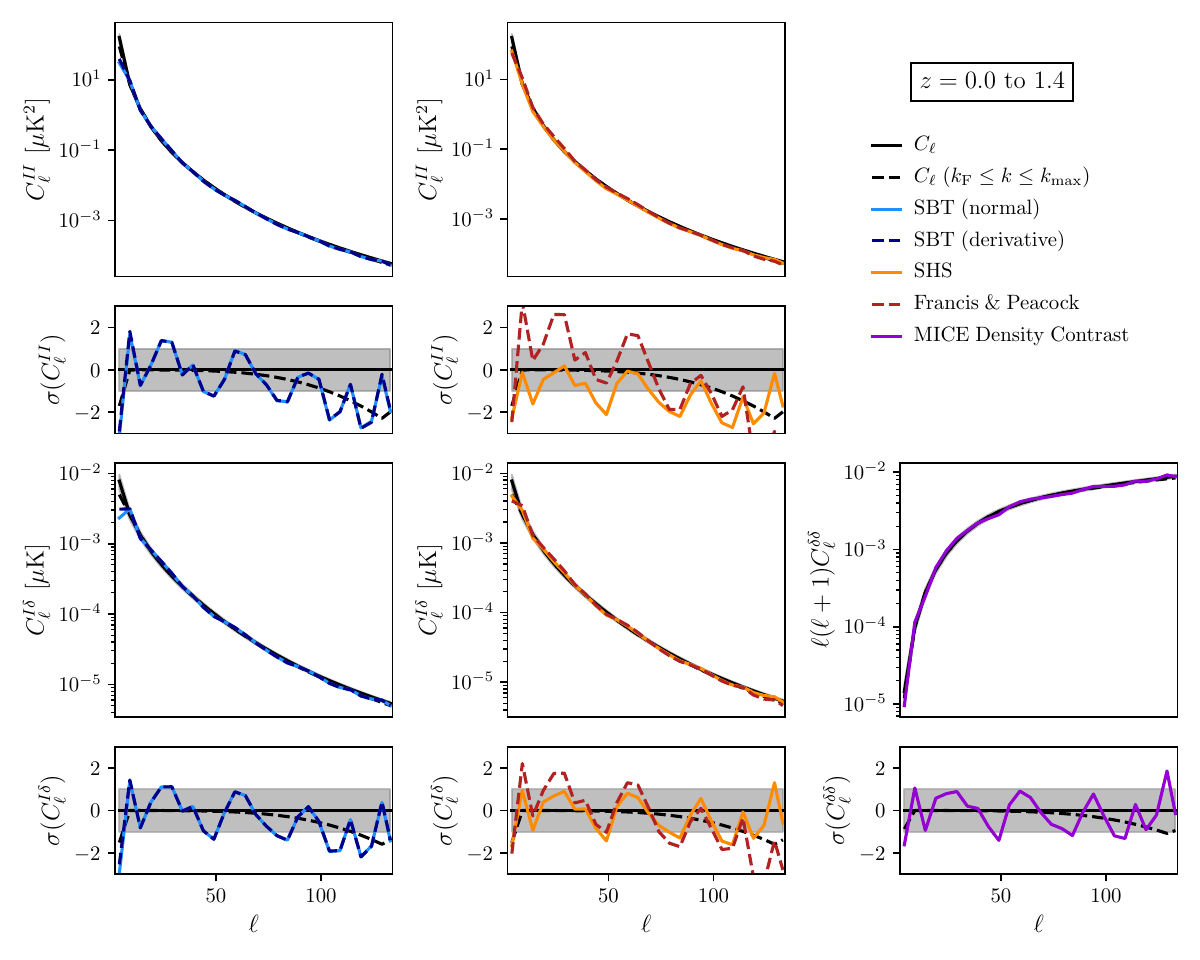}
	\caption{The auto- and cross-angular power spectra for the ISW and density field for contributions in the redshift range $0\leq z\leq 1.4$. In the top two panels we show the ISW angular power spectra $C_{\ell}^{II}$ for the SBT methods (left) and the SHS and FP approximations (middle). In the bottom panels we show the cross-angular power spectra for the ISW and density field for the SBT method (left) and the SHS and FP approximations (middle) and show the auto-angular power spectra for the MICE density field (right). In the subplots we show the significance of deviations from the full theoretical $C_{\ell}$ (i.e. with no cuts in $k$-range shown with full black lines) where $\sigma(C_{\ell})=(C_{\ell}^{\rm Measured} - C_{\ell}^{\rm Theory})/\Delta C_{\ell}$. The grey bands indicate $1\sigma$ confidence regions. We see that $C_{\ell}^{II}$ and $C_{\ell}^{I\delta}$ follow the trend of $C_{\ell} (k_{\rm F}\leq k\leq k_{\max})$ (shown with a dashed black line), i.e. beginning to fall in amplitude at high-$\ell$ with respect to the full $C_{\ell}$ mainly due to non-linear contributions at low redshift and drop at low-$\ell$ due to the lack of modes larger than the simulation box. The SHS approximation is calculated using the very $C_{\ell}$ that we compare to, so this comparison is rather redundant as we expect a reasonably good fit by construction. However in the case of the SBT and FP approximation no $C_{\ell}$ are given and this comparison represents a true test of the accuracy of these methods. The SBT ISW maps show the closest match to the theoretical $C_{\ell}$, while the approximations appear to be slightly too high in amplitude for the FP approximation and too low for the SHS approximation. Furthermore, we see the SBT methods deviate only at very low $\ell$.
	}
	\label{cls}
\end{figure*}

\section{Results: MICE Integrated Sachs-Wolfe Maps}
\label{results}

\subsection{Using Spherical Bessel Transforms}

The ISW maps for MICE are constructed using the SBT with both normal and derivative boundary conditions. We compute the SBT coefficients up to an $r_{\max}$ that corresponds to $z=2$, exceeding the maximum redshift of $z=1.4$ for which the ISW is computed. This is to ensure that no artifacts are measured near the boundary. The choice of $r_{\max}=r(z=2)$ and $k_{\max}=0.1\,h{\rm Mpc^{-1}}$ means we only need to compute the SBT up to $\ell_{\max}=380$ and $n_{\max}=121$. The full-sky MICE ISW map for contributions in the range $0\leq z\leq 1.4$ is shown in Fig. \ref{mice_isw} using the SBT with normal boundary conditions. The ISW map constructed using the SBT with derivative boundary conditions is not highlighted here, as the map is almost identical to the one shown in Fig. \ref{mice_isw}; it will be discussed in Section \ref{isw_map_comparison}.

To highlight the scales of the features that contribute at different redshifts, we plot in Fig. \ref{mice_isw_vs_z} the ISW constructed using the SBT with normal boundary conditions for redshift slices of $\Delta z=0.2$ between $z=0$ and $z=1.2$. Features in the map at low redshift are large in angular scale and become smaller at higher redshift.

\subsection{Comparison to SHS and Francis \& Peacock Approximation}
\label{isw_map_comparison}

The ISW maps for MICE have been constructed using four methods: the first two use the SBT with normal and derivative boundary conditions, while the latter two use the SHS and the FP approximation. The latter two assume no cuts in the Fourier modes used, while for the SBT the modes have been explicitly limited to $k_{\rm F}\leq k\leq k_{\max}$. To ensure the SHS and the FP approximations are comparable to those of the SBT, and do not amplify spurious Fourier modes, we multiply the spherical harmonic coefficients by $C^{II}_{\ell}(k_{\rm F}\leq k\leq k_{\max})/C_{\ell}^{II}$. Furthermore in the case of the FP approximation, since this method works best for thin shells we compute the ISW contributions in slices of $\Delta z=0.2$ which are then combined for contributions in the range $0\leq z\leq 1.4$.

In Fig. \ref{mice_comparisons_maps} the ISW maps for the four methods are shown for contributions in the range $0\leq z\leq 1.4$. In Fig. \ref{mice_comparisons_dif} the difference with respect to the SBT with normal boundary conditions is shown. These Figures show strong agreement between SBT with normal and derivative boundary conditions, but find the difference between the SHS approximation and the FP approximation to be fairly significant (deviations of the order of $\sim 2\sigma $) driven by an absence of LOS density information for the latter two methods.

\begin{figure}
	\centering
	\includegraphics[width=\columnwidth]{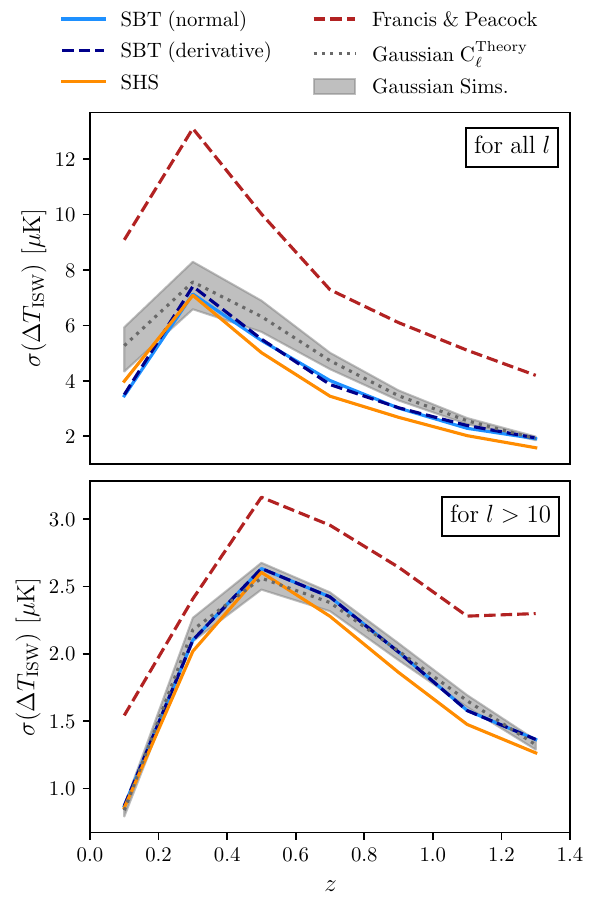}
	\caption{The standard deviation of the ISW maps for contributions in the range $0\leq z\leq 1.4$ with $\Delta z=0.2$. The four methods are compared to Gaussian realisations, where the $1\sigma$ confidence region is shown in grey. The theoretical standard deviation calculated from Eq. \ref{eq:theory_sigma_isw} is shown with grey dotted lines. In the top panel we consider the full ISW maps with no alterations and in the bottom panel we limit the maps to spherical harmonic coefficients with $\ell > 10$. The Figure shows that the FP approximation is consistently too large and the SHS slightly too low when compared to Gaussian realisations. The SBT methods are the most consistent with Gaussian realisation with the exception of some ranges in $z$ where they appear slightly lower for the full ISW map.}
	\label{isw_std_vs_z}
\end{figure}

\subsection{Angular Power Spectra: Comparisons to Theory}

All the methods in this study express the ISW in terms of its spherical harmonics. For this reason the auto- and cross-angular power spectra of the ISW and density field maps are calculated using \textsc{healpy}'s \texttt{alm2cl} function.  In Fig. \ref{cls} we plot the auto- and cross-angular power spectra for the four ISW maps and for the density field. This is compared to theoretical angular power spectra with and without cuts in the Fourier modes considered. For the full $C_{\ell}$ the integration in $k$ space is carried out for a $k$-range of $10^{-4}\leq k\leq 1\, h{\rm Mpc^{-1}}$, while the cut-$C_{\ell}$ uses a $k$-range of $k_{\rm F}\leq k\leq 0.1\, h{\rm Mpc^{-1}}$. The cut-$C_{\ell}$ are used to determine the ranges in $\ell$ for which the maps are valid. Furthermore, they indicate on which $\ell$-ranges the non-linear scales become relevant (where `non-linear' refers to small scales with $k>0.1\,h{\rm Mpc^{-1}}$). For the theoretical $C_{\ell}$ we calculate it fully without using the Limber approximation. This is because comparisons of the ISW for thin shells with $\Delta z=0.2$ showed an offset in the amplitude at all $\ell$ when compared to the measured $C_{\ell}$. The $C_{\ell}$ are binned into $\Delta \ell = 5$ and are shown in Fig. \ref{cls} to agree very well (for all methods) to the theory. The $C_{\ell}$ also reveal that the SBT ISW maps are in closest agreement to theoretical expectations, the SHS map appears to be low in power while FP appears to be slightly higher in power than would be expected (however these deviations are consistent with cosmic variance $\Delta C_{\ell}=\sqrt{2C_{\ell}^{2}/(2\ell+1)}$).

\subsection{Dependence on Redshift and Comparison to Gaussian Realisations}

In this section we measure the standard deviation of the ISW maps for contributions in the range $0\leq z\leq 1.4$ with $\Delta z=0.2$. These are compared to Gaussian realisations computed using \textsc{healpy}'s \texttt{synalm} function from the theoretical $C_{\ell}(k_{\rm F}\leq k\leq k_{\max})$. The results are shown in Fig. \ref{isw_std_vs_z} where in the top panel we consider the standard deviation of the ISW maps with no alteration and in the bottom panel we limit to only spherical harmonic coefficients $\ell>10$. The theoretical standard deviations (shown with dotted grey lines) are calculated from
\begin{equation}
  \label{eq:theory_sigma_isw}
  \sigma = \left(\sum_{\ell}\frac{2\ell + 1}{4\upi}C_{\ell}\right)^{1/2},
\end{equation}
\citep{Tegmark1997}. The comparisons show that the standard deviation in the ISW for the FP approximation is consistently too high and for the SHS is slightly low with respect to Gaussian realisations. In comparison the SBT methods are consistent with Gaussian realisations for $\ell>10$ but are low for certain redshifts when considering the full map with no alterations. The plot also indicates where the biggest contributions to the ISW occur, peaking around $z\simeq 0.3$ for all scales and at $z\simeq 0.5$ for features with $\ell>10$. This demonstrates the benefit of using the SBT methods over the SHS or FP approximation as they are better able to reproduce the theoretically expected standard deviation for the ISW temperature maps.

\subsection{Maps and Ancillary Data}

The ISW maps constructed in this study are public.\footnote{\href{https://doi.org/10.5281/zenodo.4088697}{https://doi.org/10.5281/zenodo.4088697}} They are constructed for the full redshift range $0\leq z\leq1.4$ and for spherical shells with redshift width $\Delta z=0.2$. The maps themselves are provided in \texttt{fits} format as \textsc{HEALPix} maps with \texttt{nside=256}. The spherical harmonics are also provided so that the maps can be generated to the desired \texttt{nside}. Theoretical auto- and cross-angular power spectra are provided for the ISW and density field with and without using the Limber approximation.

\section{Discussion}
\label{discussion}

In this paper we construct the ISW map for the MICE lightcone simulation and develop a pipeline for quickly constructing the ISW for future simulations; the SBT transform (the bottleneck of this computation) was calculated in $\sim 10$ minutes on a single core on commercially available hardware for $\ell_{\max}=380$ and $n_{\max}=121$. A significant computational limitation of constructing simulated ISW maps is the calculation of the time derivative of the gravitational potential $\dot{\Phi}$. Such a calculation requires the regular output of snapshots, which is memory intensive, and cannot be performed ad-hoc. To remove this obstacle we calculate the ISW maps fully in the linear regime (meaning the Poisson equation only needs to be solved once), sacrificing the construction of the non-linear ISW or Rees-Sciama effect. Since the Rees-Sciama effect at low redshift $z\lesssim 1$, at large angular scales, and after matter-domination is comparatively very small \citep[$\sim1\%$ of the ISW;][]{Seljak1996,Nadathur2014} this sacrifice has a negligible impact on the maps created. This approximation enables the ISW to be computed from the density field at $z=0$, where we insert the time evolution of the gravitational potential analytically using linear growth functions. For lightcone simulations such as MICE we divide the density field at redshift $z$ by the linear growth $D(z)$ to approximate the density field at $z=0$.

The \textsc{Python} module developed in this study, \textsc{pyGenISW}, is publicly available\footnote{\href{https://github.com/knaidoo29/pyGenISW}{https://github.com/knaidoo29/pyGenISW}} and can be used to construct the ISW using the SBT method (the focus of this paper), as well as using the SHS and the FP approximation. The ISW maps and ancillary data (spherical harmonic coefficients and theoretical angular power spectra) are made public\footnote{\href{https://doi.org/10.5281/zenodo.4088697}{https://doi.org/10.5281/zenodo.4088697}} to facilitate future LSS-CMB cross-correlation studies using MICE.

In Fig. \ref{mice_isw} we highlight one of the main data products of this study -- the ISW map for contributions in the redshift range $0\leq z\leq1.4$ for the SBT method with normal boundary conditions. In Fig. \ref{mice_isw_vs_z} the maps are constructed for redshift intervals with $\Delta z=0.2$. These maps show the redshift evolution of the scales of features and will enable isolated studies of the ISW at different redshift. We compare the SBT methods, the SHS and the FP approximation in Fig. \ref{mice_comparisons_maps} and subtract the SBT (normal) map from the other three in Fig. \ref{mice_comparisons_dif} demonstrating that the two SBT methods are virtually identical (since they differ only in their boundary conditions) and that there are significant large angular scale differences between the SBT methods and the SHS and FP approximation (driven by the absense of LOS density information). In Fig. \ref{cls} the auto- and cross-angular power spectra for the ISW and density field are shown. In these plots we see that the SBT method is in greater agreement with theoretical expectations while SHS appears to be slightly too low and the FP slightly too large at all scales, demonstrating the benefit of using the SBT methods over the SHS or FP approximation. This effect is illustrated again in Fig. \ref{isw_std_vs_z} where we compare the standard deviation of the ISW maps to those from Gaussian realisations. Once again the SBT methods are in greater agreement with theoretical expectations showing that these particular maps are the best maps to use for future MICE LSS-CMB cross-correlation studies.

These maps will be of particular relevance to large-area galaxy surveys, such as Dark Energy Survey (DES)\footnote{\href{http://www.darkenergysurvey.org}{http://www.darkenergysurvey.org}}, Dark Energy Spectroscopic Instrument (DESI)\footnote{\href{http://desi.lbl.gov/}{http://desi.lbl.gov/}}, \emph{Euclid}\footnote{\href{http://www.euclid-ec.org/}{http://www.euclid-ec.org/}}, and the Rubin Legacy Survey of Space and Time (LSST)\footnote{\href{https://www.lsst.org/}{https://www.lsst.org/}}. Cross-correlation studies have been performed on several galaxy surveys to constrain the standard cosmological model $\Lambda$CDM and to test extensions/alternative models \citep[see e.g.][]{Hang2020}. The availability of the MICE ISW maps will enable future studies to test their cross-correlation pipelines and to test the results against the predictions of the fiducial $\Lambda$CDM model. Furthermore, ISW reconstruction methods \citep{Barreiro2008, Granett2009, Manzotti2014, Muir2016} will be able to use these maps as a ground truth.

The parameters used by MICE are set according to the best-fit WMAP results and are different to the best-fit Planck results, notably in terms of $\Omega_{\rm m}$ which is measured by Planck to be $\sim 0.31$. This means the MICE ISW maps produced in this study have slightly larger ISW signals than would be expected from the best-fit Planck cosmology. However, the ISW is measured currently at low signal-to-noise meaning these differences will be negligible. Any future tests of $\Lambda$CDM from these ISW maps will need to be aware of these distinctions.

ISW void-stacking measurements are in slight tension with predictions from $\Lambda$CDM \citep[a tension of $\gtrsim2\sigma$][]{Granett2008, Papai2011, Nadathur2012, Flender2013, Hernandez2013, Illic2013, Granett2015, Kovacs2016, Nadathur2016, Cai2017, Kovacs2018, Kovacs2019}. An explanation for the source of this tension remains unclear. Recently, \citet{Kovacs2020} showed that the signal could be explained if modifications to the growth history were applied but \citet{Hang2020} show this solution is incompatible with cross-correlation studies. To determine whether variations in the value of cosmological parameters or models could explain this excess signal we will need to construct the ISW map for a large range of simulations with realistic galaxy catalogues to construct observable-like void catalogues. The pipeline developed in this study enables future work to construct the ISW from existing and future simulation suites, allowing us to understand the parameter and model dependence of the ISW void-stacking measurement. Accurate ISW maps from simulations such as MICE will enable cosmologists to fully exploit future galaxy surveys by combining probes from the early Universe in the form of the CMB and the late Universe in form of LSS. This will provide further tests of the standard cosmological model $\Lambda$CDM and may be crucial in establishing the validity of extended and alternative models. In future work we plan to provide MICE ISW maps for contributions at higher redshift and to higher $\ell_{\max}$ (essentially pushing our scale limits from the linear to the quasi-linear regime) by exploring extensions to the pipeline discussed in this paper as well as generating ISW maps for existing simulation suites probing different cosmological models and parameters.

\section*{ACKNOWLEDGEMENTS}

We thank Andras Kovacs, Peter Coles and Andrew Pontzen for providing useful comments
and discussions. KN acknowledges support from the Science and Technology Facilities
Council grant ST/N50449X and from the (Polish) National Science Centre grant \#2018/31/G/ST9/03388. PF acknowledges support from MINECO through grant ESP2017-89838-C3-1-R, the H2020 European Union grants LACEGAL 734374 and EWC 776247 with ERDF funds, and Generalitat de Catalunya through CERCA to grant 2017-SGR-885 and funding to IEEC. OL acknowledges support from an STFC Consolidated Grant ST/R000476/1.

The analysis in this paper was constructed using the following \textsc{Python} \citep{python3} modules: \textsc{NumPy} \citep{Numpy2020}, \textsc{SciPy} \citep{Scipy2020}, \textsc{healpy} \citep{Gorski2005,Zonca2019} and \textsc{camb} \citep{Lewis1999}; figures were constructed using \textsc{Matplotlib} \citep{Hunter2007} and \textsc{Cartopy} \citep{Cartopy}.

\section*{Data Availability}

The MICE density field maps used in this study can be provided upon reasonable request to the main author. The ISW maps and ancillary data are public and can be obtained from \href{https://doi.org/10.5281/zenodo.4088697}{https://doi.org/10.5281/zenodo.4088697}.



\bibliographystyle{mnras}
\bibliography{bib} 

\begin{thebibliography}{}
\makeatletter
\relax
\def\mn@urlcharsother{\let\do\@makeother \do\$\do\&\do\#\do\^\do\_\do\%\do\~}
\def\mn@doi{\begingroup\mn@urlcharsother \@ifnextchar [ {\mn@doi@}
  {\mn@doi@[]}}
\def\mn@doi@[#1]#2{\def\@tempa{#1}\ifx\@tempa\@empty \href
  {http://dx.doi.org/#2} {doi:#2}\else \href {http://dx.doi.org/#2} {#1}\fi
  \endgroup}
\def\mn@eprint#1#2{\mn@eprint@#1:#2::\@nil}
\def\mn@eprint@arXiv#1{\href {http://arxiv.org/abs/#1} {{\tt arXiv:#1}}}
\def\mn@eprint@dblp#1{\href {http://dblp.uni-trier.de/rec/bibtex/#1.xml}
  {dblp:#1}}
\def\mn@eprint@#1:#2:#3:#4\@nil{\def\@tempa {#1}\def\@tempb {#2}\def\@tempc
  {#3}\ifx \@tempc \@empty \let \@tempc \@tempb \let \@tempb \@tempa \fi \ifx
  \@tempb \@empty \def\@tempb {arXiv}\fi \@ifundefined
  {mn@eprint@\@tempb}{\@tempb:\@tempc}{\expandafter \expandafter \csname
  mn@eprint@\@tempb\endcsname \expandafter{\@tempc}}}

\bibitem[\protect\citeauthoryear{{Adamek}, {Rasera}, {Corasaniti}  \&
  {Alimi}}{{Adamek} et~al.}{2020}]{Adamek2020}
{Adamek} J.,  {Rasera} Y.,  {Corasaniti} P.~S.,   {Alimi} J.-M.,  2020, \mn@doi
  [\prd] {10.1103/PhysRevD.101.023512}, \href
  {https://ui.adsabs.harvard.edu/abs/2020PhRvD.101b3512A} {101, 023512}

\bibitem[\protect\citeauthoryear{{Afshordi}, {Loh}  \& {Strauss}}{{Afshordi}
  et~al.}{2004}]{Afshordi2004}
{Afshordi} N.,  {Loh} Y.-S.,   {Strauss} M.~A.,  2004, \mn@doi [\prd]
  {10.1103/PhysRevD.69.083524}, \href
  {https://ui.adsabs.harvard.edu/abs/2004PhRvD..69h3524A} {69, 083524}

\bibitem[\protect\citeauthoryear{{Barreiro}, {Vielva}, {Hernandez-Monteagudo}
  \& {Martinez-Gonzalez}}{{Barreiro} et~al.}{2008}]{Barreiro2008}
{Barreiro} R.~B.,  {Vielva} P.,  {Hernandez-Monteagudo} C.,
  {Martinez-Gonzalez} E.,  2008, \mn@doi [IEEE Journal of Selected Topics in
  Signal Processing] {10.1109/JSTSP.2008.2005350}, \href
  {https://ui.adsabs.harvard.edu/abs/2008ISTSP...2..747B} {2, 747}

\bibitem[\protect\citeauthoryear{{Barreiro}, {Vielva}, {Marcos-Caballero}  \&
  {Mart{\'\i}nez-Gonz{\'a}lez}}{{Barreiro} et~al.}{2013}]{Barreiro2013}
{Barreiro} R.~B.,  {Vielva} P.,  {Marcos-Caballero} A.,
  {Mart{\'\i}nez-Gonz{\'a}lez} E.,  2013, \mn@doi [\mnras]
  {10.1093/mnras/sts600}, \href
  {https://ui.adsabs.harvard.edu/abs/2013MNRAS.430..259B} {430, 259}

\bibitem[\protect\citeauthoryear{{Boughn} \& {Crittenden}}{{Boughn} \&
  {Crittenden}}{2002}]{Boughn2002}
{Boughn} S.~P.,  {Crittenden} R.~G.,  2002, \mn@doi [\prl]
  {10.1103/PhysRevLett.88.021302}, \href
  {https://ui.adsabs.harvard.edu/abs/2002PhRvL..88b1302B} {88, 021302}

\bibitem[\protect\citeauthoryear{{Cabass}, {Gerbino}, {Giusarma}, {Melchiorri},
  {Pagano}  \& {Salvati}}{{Cabass} et~al.}{2015}]{Cabass2015}
{Cabass} G.,  {Gerbino} M.,  {Giusarma} E.,  {Melchiorri} A.,  {Pagano} L.,
  {Salvati} L.,  2015, \mn@doi [\prd] {10.1103/PhysRevD.92.063534}, \href
  {https://ui.adsabs.harvard.edu/abs/2015PhRvD..92f3534C} {92, 063534}

\bibitem[\protect\citeauthoryear{{Cai}, {Cole}, {Jenkins}  \& {Frenk}}{{Cai}
  et~al.}{2009}]{Cai2009}
{Cai} Y.-C.,  {Cole} S.,  {Jenkins} A.,   {Frenk} C.,  2009, \mn@doi [\mnras]
  {10.1111/j.1365-2966.2009.14780.x}, \href
  {https://ui.adsabs.harvard.edu/abs/2009MNRAS.396..772C} {396, 772}

\bibitem[\protect\citeauthoryear{{Cai}, {Cole}, {Jenkins}  \& {Frenk}}{{Cai}
  et~al.}{2010}]{Cai2010}
{Cai} Y.-C.,  {Cole} S.,  {Jenkins} A.,   {Frenk} C.~S.,  2010, \mn@doi
  [\mnras] {10.1111/j.1365-2966.2010.16946.x}, \href
  {https://ui.adsabs.harvard.edu/abs/2010MNRAS.407..201C} {407, 201}

\bibitem[\protect\citeauthoryear{{Cai}, {Neyrinck}, {Mao}, {Peacock}, {Szapudi}
   \& {Berlind}}{{Cai} et~al.}{2017}]{Cai2017}
{Cai} Y.-C.,  {Neyrinck} M.,  {Mao} Q.,  {Peacock} J.~A.,  {Szapudi} I.,
  {Berlind} A.~A.,  2017, \mn@doi [\mnras] {10.1093/mnras/stw3299}, \href
  {https://ui.adsabs.harvard.edu/abs/2017MNRAS.466.3364C} {466, 3364}

\bibitem[\protect\citeauthoryear{{Carbone}, {Petkova}  \& {Dolag}}{{Carbone}
  et~al.}{2016}]{Carbone2016}
{Carbone} C.,  {Petkova} M.,   {Dolag} K.,  2016, \mn@doi [\jcap]
  {10.1088/1475-7516/2016/07/034}, \href
  {https://ui.adsabs.harvard.edu/abs/2016JCAP...07..034C} {2016, 034}

\bibitem[\protect\citeauthoryear{{Carretero}, {Castander}, {Gazta{\~n}aga},
  {Crocce}  \& {Fosalba}}{{Carretero} et~al.}{2015}]{MICE42015}
{Carretero} J.,  {Castander} F.~J.,  {Gazta{\~n}aga} E.,  {Crocce} M.,
  {Fosalba} P.,  2015, \mn@doi [\mnras] {10.1093/mnras/stu2402}, \href
  {https://ui.adsabs.harvard.edu/abs/2015MNRAS.447..646C} {447, 646}

\bibitem[\protect\citeauthoryear{{Carretero} et~al.,}{{Carretero}
  et~al.}{2017}]{Cosmohub1}
{Carretero} J.,  et~al., 2017, in Proceedings of the European Physical Society
  Conference on High Energy Physics. 5-12 July. p.~488

\bibitem[\protect\citeauthoryear{{Corasaniti}, {Giannantonio}  \&
  {Melchiorri}}{{Corasaniti} et~al.}{2005}]{Corasaniti2005}
{Corasaniti} P.-S.,  {Giannantonio} T.,   {Melchiorri} A.,  2005, \mn@doi
  [\prd] {10.1103/PhysRevD.71.123521}, \href
  {https://ui.adsabs.harvard.edu/abs/2005PhRvD..71l3521C} {71, 123521}

\bibitem[\protect\citeauthoryear{{Crittenden} \& {Turok}}{{Crittenden} \&
  {Turok}}{1996}]{Crittenden1996}
{Crittenden} R.~G.,  {Turok} N.,  1996, \mn@doi [\prl]
  {10.1103/PhysRevLett.76.575}, \href
  {https://ui.adsabs.harvard.edu/abs/1996PhRvL..76..575C} {76, 575}

\bibitem[\protect\citeauthoryear{{Crocce}, {Castander}, {Gazta{\~n}aga},
  {Fosalba}  \& {Carretero}}{{Crocce} et~al.}{2015}]{MICE22015}
{Crocce} M.,  {Castander} F.~J.,  {Gazta{\~n}aga} E.,  {Fosalba} P.,
  {Carretero} J.,  2015, \mn@doi [\mnras] {10.1093/mnras/stv1708}, \href
  {https://ui.adsabs.harvard.edu/abs/2015MNRAS.453.1513C} {453, 1513}

\bibitem[\protect\citeauthoryear{{Dunkley} et~al.,}{{Dunkley}
  et~al.}{2009}]{Dunkley2009}
{Dunkley} J.,  et~al., 2009, \mn@doi [\apjs] {10.1088/0067-0049/180/2/306},
  \href {https://ui.adsabs.harvard.edu/abs/2009ApJS..180..306D} {180, 306}

\bibitem[\protect\citeauthoryear{{Dup{\'e}}, {Rassat}, {Starck}  \&
  {Fadili}}{{Dup{\'e}} et~al.}{2011}]{Dupe2011}
{Dup{\'e}} F.~X.,  {Rassat} A.,  {Starck} J.~L.,   {Fadili} M.~J.,  2011,
  \mn@doi [\aap] {10.1051/0004-6361/201015893}, \href
  {https://ui.adsabs.harvard.edu/abs/2011A&A...534A..51D} {534, A51}

\bibitem[\protect\citeauthoryear{{Ferraro}, {Sherwin}  \& {Spergel}}{{Ferraro}
  et~al.}{2015}]{Ferraro2015}
{Ferraro} S.,  {Sherwin} B.~D.,   {Spergel} D.~N.,  2015, \mn@doi [\prd]
  {10.1103/PhysRevD.91.083533}, \href
  {https://ui.adsabs.harvard.edu/abs/2015PhRvD..91h3533F} {91, 083533}

\bibitem[\protect\citeauthoryear{{Flender}, {Hotchkiss}  \&
  {Nadathur}}{{Flender} et~al.}{2013}]{Flender2013}
{Flender} S.,  {Hotchkiss} S.,   {Nadathur} S.,  2013, \mn@doi [\jcap]
  {10.1088/1475-7516/2013/02/013}, \href
  {https://ui.adsabs.harvard.edu/abs/2013JCAP...02..013F} {2013, 013}

\bibitem[\protect\citeauthoryear{{Fosalba}, {Gazta{\~n}aga}  \&
  {Castander}}{{Fosalba} et~al.}{2003}]{Fosalba2003}
{Fosalba} P.,  {Gazta{\~n}aga} E.,   {Castander} F.~J.,  2003, \mn@doi [\apjl]
  {10.1086/379848}, \href
  {https://ui.adsabs.harvard.edu/abs/2003ApJ...597L..89F} {597, L89}

\bibitem[\protect\citeauthoryear{{Fosalba}, {Gazta{\~n}aga}, {Castander}  \&
  {Manera}}{{Fosalba} et~al.}{2008}]{Fosalba2008}
{Fosalba} P.,  {Gazta{\~n}aga} E.,  {Castander} F.~J.,   {Manera} M.,  2008,
  \mn@doi [\mnras] {10.1111/j.1365-2966.2008.13910.x}, \href
  {https://ui.adsabs.harvard.edu/abs/2008MNRAS.391..435F} {391, 435}

\bibitem[\protect\citeauthoryear{{Fosalba}, {Gazta{\~n}aga}, {Castander}  \&
  {Crocce}}{{Fosalba} et~al.}{2015a}]{MICE32015}
{Fosalba} P.,  {Gazta{\~n}aga} E.,  {Castander} F.~J.,   {Crocce} M.,  2015a,
  \mn@doi [\mnras] {10.1093/mnras/stu2464}, \href
  {https://ui.adsabs.harvard.edu/abs/2015MNRAS.447.1319F} {447, 1319}

\bibitem[\protect\citeauthoryear{{Fosalba}, {Crocce}, {Gazta{\~n}aga}  \&
  {Castander}}{{Fosalba} et~al.}{2015b}]{MICE2015}
{Fosalba} P.,  {Crocce} M.,  {Gazta{\~n}aga} E.,   {Castander} F.~J.,  2015b,
  \mn@doi [\mnras] {10.1093/mnras/stv138}, \href
  {https://ui.adsabs.harvard.edu/abs/2015MNRAS.448.2987F} {448, 2987}

\bibitem[\protect\citeauthoryear{{Francis} \& {Peacock}}{{Francis} \&
  {Peacock}}{2010}]{Francis2010}
{Francis} C.~L.,  {Peacock} J.~A.,  2010, \mn@doi [\mnras]
  {10.1111/j.1365-2966.2010.16278.x}, \href
  {https://ui.adsabs.harvard.edu/abs/2010MNRAS.406....2F} {406, 2}

\bibitem[\protect\citeauthoryear{{Giannantonio} et~al.,}{{Giannantonio}
  et~al.}{2006}]{giannantoni2006}
{Giannantonio} T.,  et~al., 2006, \mn@doi [\prd] {10.1103/PhysRevD.74.063520},
  \href {https://ui.adsabs.harvard.edu/abs/2006PhRvD..74f3520G} {74, 063520}

\bibitem[\protect\citeauthoryear{{Giannantonio}, {Scranton}, {Crittenden},
  {Nichol}, {Boughn}, {Myers}  \& {Richards}}{{Giannantonio}
  et~al.}{2008}]{giannantonio2008}
{Giannantonio} T.,  {Scranton} R.,  {Crittenden} R.~G.,  {Nichol} R.~C.,
  {Boughn} S.~P.,  {Myers} A.~D.,   {Richards} G.~T.,  2008, \mn@doi [\prd]
  {10.1103/PhysRevD.77.123520}, \href
  {https://ui.adsabs.harvard.edu/abs/2008PhRvD..77l3520G} {77, 123520}

\bibitem[\protect\citeauthoryear{{Giannantonio}, {Crittenden}, {Nichol}  \&
  {Ross}}{{Giannantonio} et~al.}{2012}]{giannantonio2012}
{Giannantonio} T.,  {Crittenden} R.,  {Nichol} R.,   {Ross} A.~J.,  2012,
  \mn@doi [\mnras] {10.1111/j.1365-2966.2012.21896.x}, \href
  {https://ui.adsabs.harvard.edu/abs/2012MNRAS.426.2581G} {426, 2581}

\bibitem[\protect\citeauthoryear{{Giannantonio}, {Ross}, {Percival},
  {Crittenden}, {Bacher}, {Kilbinger}, {Nichol}  \& {Weller}}{{Giannantonio}
  et~al.}{2014}]{giannantonio2014}
{Giannantonio} T.,  {Ross} A.~J.,  {Percival} W.~J.,  {Crittenden} R.,
  {Bacher} D.,  {Kilbinger} M.,  {Nichol} R.,   {Weller} J.,  2014, \mn@doi
  [\prd] {10.1103/PhysRevD.89.023511}, \href
  {https://ui.adsabs.harvard.edu/abs/2014PhRvD..89b3511G} {89, 023511}

\bibitem[\protect\citeauthoryear{{G{\'o}rski}, {Hivon}, {Banday}, {Wandelt},
  {Hansen}, {Reinecke}  \& {Bartelmann}}{{G{\'o}rski}
  et~al.}{2005}]{Gorski2005}
{G{\'o}rski} K.~M.,  {Hivon} E.,  {Banday} A.~J.,  {Wandelt} B.~D.,  {Hansen}
  F.~K.,  {Reinecke} M.,   {Bartelmann} M.,  2005, \mn@doi [\apj]
  {10.1086/427976}, \href
  {https://ui.adsabs.harvard.edu/abs/2005ApJ...622..759G} {622, 759}

\bibitem[\protect\citeauthoryear{{Goto}, {Szapudi}  \& {Granett}}{{Goto}
  et~al.}{2012}]{Tomotsugu2012}
{Goto} T.,  {Szapudi} I.,   {Granett} B.~R.,  2012, \mn@doi [\mnras]
  {10.1111/j.1745-3933.2012.01240.x}, \href
  {https://ui.adsabs.harvard.edu/abs/2012MNRAS.422L..77G} {422, L77}

\bibitem[\protect\citeauthoryear{{Granett}, {Neyrinck}  \& {Szapudi}}{{Granett}
  et~al.}{2008}]{Granett2008}
{Granett} B.~R.,  {Neyrinck} M.~C.,   {Szapudi} I.,  2008, \mn@doi [\apjl]
  {10.1086/591670}, \href
  {https://ui.adsabs.harvard.edu/abs/2008ApJ...683L..99G} {683, L99}

\bibitem[\protect\citeauthoryear{{Granett}, {Neyrinck}  \& {Szapudi}}{{Granett}
  et~al.}{2009}]{Granett2009}
{Granett} B.~R.,  {Neyrinck} M.~C.,   {Szapudi} I.,  2009, \mn@doi [\apj]
  {10.1088/0004-637X/701/1/414}, \href
  {https://ui.adsabs.harvard.edu/abs/2009ApJ...701..414G} {701, 414}

\bibitem[\protect\citeauthoryear{{Granett}, {Kov{\'a}cs}  \&
  {Hawken}}{{Granett} et~al.}{2015}]{Granett2015}
{Granett} B.~R.,  {Kov{\'a}cs} A.,   {Hawken} A.~J.,  2015, \mn@doi [\mnras]
  {10.1093/mnras/stv2110}, \href
  {https://ui.adsabs.harvard.edu/abs/2015MNRAS.454.2804G} {454, 2804}

\bibitem[\protect\citeauthoryear{{Hang}, {Alam}, {Peacock}  \& {Cai}}{{Hang}
  et~al.}{2020}]{Hang2020}
{Hang} Q.,  {Alam} S.,  {Peacock} J.~A.,   {Cai} Y.-C.,  2020, arXiv e-prints,
  \href {https://ui.adsabs.harvard.edu/abs/2020arXiv201000466H} {p.
  arXiv:2010.00466}

\bibitem[\protect\citeauthoryear{Harris et~al.,}{Harris
  et~al.}{2020}]{Numpy2020}
Harris C.~R.,  et~al., 2020, \mn@doi [Nature] {10.1038/s41586-020-2649-2}, 585,
  357

\bibitem[\protect\citeauthoryear{{Hern{\'a}ndez-Monteagudo} \&
  {Smith}}{{Hern{\'a}ndez-Monteagudo} \& {Smith}}{2013}]{Hernandez2013}
{Hern{\'a}ndez-Monteagudo} C.,  {Smith} R.~E.,  2013, \mn@doi [\mnras]
  {10.1093/mnras/stt1322}, \href
  {https://ui.adsabs.harvard.edu/abs/2013MNRAS.435.1094H} {435, 1094}

\bibitem[\protect\citeauthoryear{{Ho}, {Hirata}, {Padmanabhan}, {Seljak}  \&
  {Bahcall}}{{Ho} et~al.}{2008}]{Ho2008}
{Ho} S.,  {Hirata} C.,  {Padmanabhan} N.,  {Seljak} U.,   {Bahcall} N.,  2008,
  \mn@doi [\prd] {10.1103/PhysRevD.78.043519}, \href
  {https://ui.adsabs.harvard.edu/abs/2008PhRvD..78d3519H} {78, 043519}

\bibitem[\protect\citeauthoryear{{Hoffmann}, {Bel}, {Gazta{\~n}aga}, {Crocce},
  {Fosalba}  \& {Castander}}{{Hoffmann} et~al.}{2015}]{MICE52015}
{Hoffmann} K.,  {Bel} J.,  {Gazta{\~n}aga} E.,  {Crocce} M.,  {Fosalba} P.,
  {Castander} F.~J.,  2015, \mn@doi [\mnras] {10.1093/mnras/stu2492}, \href
  {https://ui.adsabs.harvard.edu/abs/2015MNRAS.447.1724H} {447, 1724}

\bibitem[\protect\citeauthoryear{Hunter}{Hunter}{2007}]{Hunter2007}
Hunter J.~D.,  2007, \mn@doi [Computing in Science \& Engineering]
  {10.1109/MCSE.2007.55}, 9, 90

\bibitem[\protect\citeauthoryear{{Ili{\'c}}, {Langer}  \& {Douspis}}{{Ili{\'c}}
  et~al.}{2013}]{Illic2013}
{Ili{\'c}} S.,  {Langer} M.,   {Douspis} M.,  2013, \mn@doi [\aap]
  {10.1051/0004-6361/201321150}, \href
  {https://ui.adsabs.harvard.edu/abs/2013A&A...556A..51I} {556, A51}

\bibitem[\protect\citeauthoryear{{Kov{\'a}cs}}{{Kov{\'a}cs}}{2018}]{Kovacs2018}
{Kov{\'a}cs} A.,  2018, \mn@doi [\mnras] {10.1093/mnras/stx3213}, \href
  {https://ui.adsabs.harvard.edu/abs/2018MNRAS.475.1777K} {475, 1777}

\bibitem[\protect\citeauthoryear{{Kov{\'a}cs} \&
  {Garc{\'\i}a-Bellido}}{{Kov{\'a}cs} \&
  {Garc{\'\i}a-Bellido}}{2016}]{Kovacs2016}
{Kov{\'a}cs} A.,  {Garc{\'\i}a-Bellido} J.,  2016, \mn@doi [\mnras]
  {10.1093/mnras/stw1752}, \href
  {https://ui.adsabs.harvard.edu/abs/2016MNRAS.462.1882K} {462, 1882}

\bibitem[\protect\citeauthoryear{{Kov{\'a}cs} et~al.,}{{Kov{\'a}cs}
  et~al.}{2019}]{Kovacs2019}
{Kov{\'a}cs} A.,  et~al., 2019, \mn@doi [\mnras] {10.1093/mnras/stz341}, \href
  {https://ui.adsabs.harvard.edu/abs/2019MNRAS.484.5267K} {484, 5267}

\bibitem[\protect\citeauthoryear{{Kov{\'a}cs}, {Beck}, {Szapudi}, {Csabai},
  {R{\'a}cz}  \& {Dobos}}{{Kov{\'a}cs} et~al.}{2020}]{Kovacs2020}
{Kov{\'a}cs} A.,  {Beck} R.,  {Szapudi} I.,  {Csabai} I.,  {R{\'a}cz} G.,
  {Dobos} L.,  2020, \mn@doi [\mnras] {10.1093/mnras/staa2631}, \href
  {https://ui.adsabs.harvard.edu/abs/2020MNRAS.499..320K} {499, 320}

\bibitem[\protect\citeauthoryear{Lahav, Lilje, Primack  \& Rees}{Lahav
  et~al.}{1991}]{ofer1991}
Lahav O.,  Lilje P.~B.,  Primack J.~R.,   Rees M.~J.,  1991, \mn@doi [Monthly
  Notices of the Royal Astronomical Society] {10.1093/mnras/251.1.128}, 251,
  128

\bibitem[\protect\citeauthoryear{{Leistedt}, {Rassat}, {R{\'e}fr{\'e}gier}  \&
  {Starck}}{{Leistedt} et~al.}{2012}]{Leistedt3DEX}
{Leistedt} B.,  {Rassat} A.,  {R{\'e}fr{\'e}gier} A.,   {Starck} J.~L.,  2012,
  \mn@doi [\aap] {10.1051/0004-6361/201118463}, \href
  {https://ui.adsabs.harvard.edu/abs/2012A&A...540A..60L} {540, A60}

\bibitem[\protect\citeauthoryear{Lewis, Challinor  \& Lasenby}{Lewis
  et~al.}{2000}]{Lewis1999}
Lewis A.,  Challinor A.,   Lasenby A.,  2000, \mn@doi [\apj] {10.1086/309179},
  538, 473

\bibitem[\protect\citeauthoryear{{Limber}}{{Limber}}{1954}]{Limber1954}
{Limber} D.~N.,  1954, \mn@doi [\apj] {10.1086/145870}, \href
  {https://ui.adsabs.harvard.edu/abs/1954ApJ...119..655L} {119, 655}

\bibitem[\protect\citeauthoryear{{Mackenzie}, {Shanks}, {Bremer}, {Cai},
  {Gunawardhana}, {Kov{\'a}cs}, {Norberg}  \& {Szapudi}}{{Mackenzie}
  et~al.}{2017}]{Mackenzie2017}
{Mackenzie} R.,  {Shanks} T.,  {Bremer} M.~N.,  {Cai} Y.-C.,  {Gunawardhana} M.
  L.~P.,  {Kov{\'a}cs} A.,  {Norberg} P.,   {Szapudi} I.,  2017, \mn@doi
  [\mnras] {10.1093/mnras/stx931}, \href
  {https://ui.adsabs.harvard.edu/abs/2017MNRAS.470.2328M} {470, 2328}

\bibitem[\protect\citeauthoryear{{Manzotti} \& {Dodelson}}{{Manzotti} \&
  {Dodelson}}{2014}]{Manzotti2014}
{Manzotti} A.,  {Dodelson} S.,  2014, \mn@doi [\prd]
  {10.1103/PhysRevD.90.123009}, \href
  {https://ui.adsabs.harvard.edu/abs/2014PhRvD..90l3009M} {90, 123009}

\bibitem[\protect\citeauthoryear{{McEwen}, {Vielva}, {Hobson},
  {Mart{\'\i}nez-Gonz{\'a}lez}  \& {Lasenby}}{{McEwen}
  et~al.}{2007}]{McEwen2007}
{McEwen} J.~D.,  {Vielva} P.,  {Hobson} M.~P.,  {Mart{\'\i}nez-Gonz{\'a}lez}
  E.,   {Lasenby} A.~N.,  2007, \mn@doi [\mnras]
  {10.1111/j.1365-2966.2007.11505.x}, \href
  {https://ui.adsabs.harvard.edu/abs/2007MNRAS.376.1211M} {376, 1211}

\bibitem[\protect\citeauthoryear{{Met Office}}{{Met Office}}{2015}]{Cartopy}
{Met Office} 2010 - 2015, Cartopy: a cartographic python library with a
  Matplotlib interface.
Exeter, Devon, \url {https://scitools.org.uk/cartopy}

\bibitem[\protect\citeauthoryear{{Muir} \& {Huterer}}{{Muir} \&
  {Huterer}}{2016}]{Muir2016}
{Muir} J.,  {Huterer} D.,  2016, \mn@doi [\prd] {10.1103/PhysRevD.94.043503},
  \href {https://ui.adsabs.harvard.edu/abs/2016PhRvD..94d3503M} {94, 043503}

\bibitem[\protect\citeauthoryear{{Nadathur} \& {Crittenden}}{{Nadathur} \&
  {Crittenden}}{2016}]{Nadathur2016}
{Nadathur} S.,  {Crittenden} R.,  2016, \mn@doi [\apjl]
  {10.3847/2041-8205/830/1/L19}, \href
  {https://ui.adsabs.harvard.edu/abs/2016ApJ...830L..19N} {830, L19}

\bibitem[\protect\citeauthoryear{{Nadathur}, {Hotchkiss}  \&
  {Sarkar}}{{Nadathur} et~al.}{2012}]{Nadathur2012}
{Nadathur} S.,  {Hotchkiss} S.,   {Sarkar} S.,  2012, \mn@doi [\jcap]
  {10.1088/1475-7516/2012/06/042}, \href
  {https://ui.adsabs.harvard.edu/abs/2012JCAP...06..042N} {2012, 042}

\bibitem[\protect\citeauthoryear{{Nadathur}, {Lavinto}, {Hotchkiss}  \&
  {R{\"a}s{\"a}nen}}{{Nadathur} et~al.}{2014}]{Nadathur2014}
{Nadathur} S.,  {Lavinto} M.,  {Hotchkiss} S.,   {R{\"a}s{\"a}nen} S.,  2014,
  \mn@doi [\prd] {10.1103/PhysRevD.90.103510}, \href
  {https://ui.adsabs.harvard.edu/abs/2014PhRvD..90j3510N} {90, 103510}

\bibitem[\protect\citeauthoryear{{Naidoo}, {Benoit-L{\'e}vy}  \&
  {Lahav}}{{Naidoo} et~al.}{2016}]{Naidoo2016}
{Naidoo} K.,  {Benoit-L{\'e}vy} A.,   {Lahav} O.,  2016, \mn@doi [\mnras]
  {10.1093/mnrasl/slw043}, \href
  {https://ui.adsabs.harvard.edu/abs/2016MNRAS.459L..71N} {459, L71}

\bibitem[\protect\citeauthoryear{{Naidoo}, {Benoit-L{\'e}vy}  \&
  {Lahav}}{{Naidoo} et~al.}{2017}]{Naidoo2017}
{Naidoo} K.,  {Benoit-L{\'e}vy} A.,   {Lahav} O.,  2017, \mn@doi [\mnras]
  {10.1093/mnrasl/slx140}, \href
  {https://ui.adsabs.harvard.edu/abs/2017MNRAS.472L..65N} {472, L65}

\bibitem[\protect\citeauthoryear{{Nicola}, {Refregier}  \& {Amara}}{{Nicola}
  et~al.}{2016}]{Nicola2016}
{Nicola} A.,  {Refregier} A.,   {Amara} A.,  2016, \mn@doi [\prd]
  {10.1103/PhysRevD.94.083517}, \href
  {https://ui.adsabs.harvard.edu/abs/2016PhRvD..94h3517N} {94, 083517}

\bibitem[\protect\citeauthoryear{{Nolta} et~al.,}{{Nolta}
  et~al.}{2004}]{Nolta2004}
{Nolta} M.~R.,  et~al., 2004, \mn@doi [\apj] {10.1086/386536}, \href
  {https://ui.adsabs.harvard.edu/abs/2004ApJ...608...10N} {608, 10}

\bibitem[\protect\citeauthoryear{{Padmanabhan}, {Hirata}, {Seljak}, {Schlegel},
  {Brinkmann}  \& {Schneider}}{{Padmanabhan} et~al.}{2005}]{Padmanabhan2005}
{Padmanabhan} N.,  {Hirata} C.~M.,  {Seljak} U.,  {Schlegel} D.~J.,
  {Brinkmann} J.,   {Schneider} D.~P.,  2005, \mn@doi [\prd]
  {10.1103/PhysRevD.72.043525}, \href
  {https://ui.adsabs.harvard.edu/abs/2005PhRvD..72d3525P} {72, 043525}

\bibitem[\protect\citeauthoryear{{P{\'a}pai}, {Szapudi}  \&
  {Granett}}{{P{\'a}pai} et~al.}{2011}]{Papai2011}
{P{\'a}pai} P.,  {Szapudi} I.,   {Granett} B.~R.,  2011, \mn@doi [\apj]
  {10.1088/0004-637X/732/1/27}, \href
  {https://ui.adsabs.harvard.edu/abs/2011ApJ...732...27P} {732, 27}

\bibitem[\protect\citeauthoryear{{Peebles}}{{Peebles}}{1980}]{Peebles1980}
{Peebles} P.~J.~E.,  1980, {The large-scale structure of the universe}.
Princeton University Press Princeton, N.J

\bibitem[\protect\citeauthoryear{{Peiris} \& {Spergel}}{{Peiris} \&
  {Spergel}}{2000}]{Peiris2000}
{Peiris} H.~V.,  {Spergel} D.~N.,  2000, \mn@doi [\apj] {10.1086/309373}, \href
  {https://ui.adsabs.harvard.edu/abs/2000ApJ...540..605P} {540, 605}

\bibitem[\protect\citeauthoryear{{Planck Collaboration} et~al.,}{{Planck
  Collaboration} et~al.}{2014}]{PlanckISW2014}
{Planck Collaboration} et~al., 2014, \mn@doi [\aap]
  {10.1051/0004-6361/201321526}, \href
  {https://ui.adsabs.harvard.edu/abs/2014A&A...571A..19P} {571, A19}

\bibitem[\protect\citeauthoryear{{Planck Collaboration} et~al.,}{{Planck
  Collaboration} et~al.}{2016}]{PlanckISW2016}
{Planck Collaboration} et~al., 2016, \mn@doi [\aap]
  {10.1051/0004-6361/201525831}, \href
  {https://ui.adsabs.harvard.edu/abs/2016A&A...594A..21P} {594, A21}

\bibitem[\protect\citeauthoryear{{Rees} \& {Sciama}}{{Rees} \&
  {Sciama}}{1968}]{Rees1968}
{Rees} M.~J.,  {Sciama} D.~W.,  1968, \mn@doi [\nat] {10.1038/217511a0}, \href
  {http://adsabs.harvard.edu/abs/1968Natur.217..511R} {217, 511}

\bibitem[\protect\citeauthoryear{{Sachs} \& {Wolfe}}{{Sachs} \&
  {Wolfe}}{1967}]{Sachs1967}
{Sachs} R.~K.,  {Wolfe} A.~M.,  1967, \mn@doi [\apj] {10.1086/148982}, \href
  {http://adsabs.harvard.edu/abs/1967ApJ...147...73S} {147, 73}

\bibitem[\protect\citeauthoryear{{Scranton} et~al.,}{{Scranton}
  et~al.}{2003}]{Scranton2003}
{Scranton} R.,  et~al., 2003, arXiv e-prints, \href
  {https://ui.adsabs.harvard.edu/abs/2003astro.ph..7335S} {pp
  astro--ph/0307335}

\bibitem[\protect\citeauthoryear{{Seljak}}{{Seljak}}{1996}]{Seljak1996}
{Seljak} U.,  1996, \mn@doi [\apj] {10.1086/176991}, \href
  {https://ui.adsabs.harvard.edu/abs/1996ApJ...460..549S} {460, 549}

\bibitem[\protect\citeauthoryear{{Shapiro}, {Crittenden}  \&
  {Percival}}{{Shapiro} et~al.}{2012}]{Shapiro2012}
{Shapiro} C.,  {Crittenden} R.~G.,   {Percival} W.~J.,  2012, \mn@doi [\mnras]
  {10.1111/j.1365-2966.2012.20785.x}, \href
  {https://ui.adsabs.harvard.edu/abs/2012MNRAS.422.2341S} {422, 2341}

\bibitem[\protect\citeauthoryear{{Springel}}{{Springel}}{2005}]{Springel2005}
{Springel} V.,  2005, \mn@doi [\mnras] {10.1111/j.1365-2966.2005.09655.x},
  \href {https://ui.adsabs.harvard.edu/abs/2005MNRAS.364.1105S} {364, 1105}

\bibitem[\protect\citeauthoryear{{St{\"o}lzner}, {Cuoco}, {Lesgourgues}  \&
  {Bilicki}}{{St{\"o}lzner} et~al.}{2018}]{Stolzner2018}
{St{\"o}lzner} B.,  {Cuoco} A.,  {Lesgourgues} J.,   {Bilicki} M.,  2018,
  \mn@doi [\prd] {10.1103/PhysRevD.97.063506}, \href
  {https://ui.adsabs.harvard.edu/abs/2018PhRvD..97f3506S} {97, 063506}

\bibitem[\protect\citeauthoryear{{Szapudi} et~al.,}{{Szapudi}
  et~al.}{2015}]{Szapudi2015}
{Szapudi} I.,  et~al., 2015, \mn@doi [\mnras] {10.1093/mnras/stv488}, \href
  {https://ui.adsabs.harvard.edu/abs/2015MNRAS.450..288S} {450, 288}

\bibitem[\protect\citeauthoryear{{Tallada} et~al.,}{{Tallada}
  et~al.}{2020}]{Cosmohub2}
{Tallada} P.,  et~al., 2020, \mn@doi [Astronomy and Computing]
  {10.1016/j.ascom.2020.100391}, \href
  {https://ui.adsabs.harvard.edu/abs/2020A&C....3200391T} {32, 100391}

\bibitem[\protect\citeauthoryear{{Tegmark}}{{Tegmark}}{1997}]{Tegmark1997}
{Tegmark} M.,  1997, \mn@doi [\prd] {10.1103/PhysRevD.56.4514}, \href
  {https://ui.adsabs.harvard.edu/abs/1997PhRvD..56.4514T} {56, 4514}

\bibitem[\protect\citeauthoryear{Van~Rossum \& Drake}{Van~Rossum \&
  Drake}{2009}]{python3}
Van~Rossum G.,  Drake F.~L.,  2009, Python 3 Reference Manual.
CreateSpace, Scotts Valley, CA

\bibitem[\protect\citeauthoryear{{Vielva}, {Mart{\'\i}nez-Gonz{\'a}lez}  \&
  {Tucci}}{{Vielva} et~al.}{2006}]{Vielva2006}
{Vielva} P.,  {Mart{\'\i}nez-Gonz{\'a}lez} E.,   {Tucci} M.,  2006, \mn@doi
  [\mnras] {10.1111/j.1365-2966.2005.09764.x}, \href
  {https://ui.adsabs.harvard.edu/abs/2006MNRAS.365..891V} {365, 891}

\bibitem[\protect\citeauthoryear{Virtanen et~al.,}{Virtanen
  et~al.}{2020}]{Scipy2020}
Virtanen P.,  et~al., 2020, \mn@doi [Nature Methods]
  {10.1038/s41592-019-0686-2}, \href {https://rdcu.be/b08Wh} {17, 261}

\bibitem[\protect\citeauthoryear{{Wang}, {Ronneberger}  \& {Burkhardt}}{{Wang}
  et~al.}{2009}]{Wang2009}
{Wang} Q.,  {Ronneberger} O.,   {Burkhardt} H.,  2009, \mn@doi [IEEE
  Transactions on Pattern Analysis and Machine Intelligence]
  {10.1109/TPAMI.2009.29}, 31, 1715

\bibitem[\protect\citeauthoryear{{Watson} et~al.,}{{Watson}
  et~al.}{2014}]{Watson2014}
{Watson} W.~A.,  et~al., 2014, \mn@doi [\mnras] {10.1093/mnras/stt2208}, \href
  {https://ui.adsabs.harvard.edu/abs/2014MNRAS.438..412W} {438, 412}

\bibitem[\protect\citeauthoryear{{Xia}, {Viel}, {Baccigalupi}  \&
  {Matarrese}}{{Xia} et~al.}{2009}]{Xia2009}
{Xia} J.-Q.,  {Viel} M.,  {Baccigalupi} C.,   {Matarrese} S.,  2009, \mn@doi
  [\jcap] {10.1088/1475-7516/2009/09/003}, \href
  {https://ui.adsabs.harvard.edu/abs/2009JCAP...09..003X} {2009, 003}

\bibitem[\protect\citeauthoryear{Zonca, Singer, Lenz, Reinecke, Rosset, Hivon
  \& Gorski}{Zonca et~al.}{2019}]{Zonca2019}
Zonca A.,  Singer L.,  Lenz D.,  Reinecke M.,  Rosset C.,  Hivon E.,   Gorski
  K.,  2019, \mn@doi [Journal of Open Source Software] {10.21105/joss.01298},
  4, 1298

\makeatother
\end{thebibliography}







\bsp	
\label{lastpage}
\end{document}